\documentclass[iop]{emulateapj}
\usepackage{appendix}

\begin{document}

\newcommand{\mirlum}{L_{\rm 8}}
\newcommand{\ebmv}{E(B-V)}
\newcommand{\lha}{L(H\alpha)}
\newcommand{\lir}{L_{\rm IR}}
\newcommand{\lbol}{L_{\rm bol}}
\newcommand{\luv}{L_{\rm UV}}
\newcommand{\rs}{{\cal R}}
\newcommand{\ugr}{U_{\rm n}G\rs}
\newcommand{\ks}{K_{\rm s}}
\newcommand{\gmr}{G-\rs}
\newcommand{\hi}{\text{\ion{H}{1}}}
\newcommand{\nhi}{N(\text{\ion{H}{1}})}
\newcommand{\lognhi}{\log[\nhi/{\rm cm}^{-2}]}
\newcommand{\molh}{\text{H}_2}
\newcommand{\nmolh}{N(\molh)}
\newcommand{\lognmolh}{\log[\nmolh/{\rm cm}^{-2}]}

\title{THE CONNECTION BETWEEN REDDENING, GAS COVERING FRACTION, AND
  THE ESCAPE OF IONIZING RADIATION AT HIGH REDSHIFT
%
\altaffilmark{*}}
\author{\sc Naveen A. Reddy\altaffilmark{1,7}, 
Charles C. Steidel\altaffilmark{2},
Max Pettini\altaffilmark{3},
Milan Bogosavljevi\'c\altaffilmark{4,5}, \&
Alice E. Shapley\altaffilmark{6}}


\altaffiltext{*}{Based on data obtained at the W.M. Keck Observatory,
  which is operated as a scientific partnership among the California
  Institute of Technology, the University of California, and NASA, and
  was made possible by the generous financial support of the W.M. Keck
  Foundation.}

\altaffiltext{1}{Department of Physics and Astronomy, University of California, 
Riverside, 900 University Avenue, Riverside, CA 92521, USA}
\altaffiltext{2}{Department of Astronomy, California Institute of Technology, 
MS 105--24, Pasadena,CA 91125, USA}
\altaffiltext{3}{Institute of Astronomy, Madingley Road, Cambridge
CB3 OHA, UK}
\altaffiltext{4}{Astronomical Observatory, Belgrade, Volgina 7, 11060 Belgrade, Serbia}
\altaffiltext{5}{New York University Abu Dhabi, P.O. Box 129188, Abu Dhabi, UAE}
\altaffiltext{6}{Department of Physics \& Astronomy, University of California,
Los Angeles, 430 Portola Plaza, Los Angeles, CA 90095}
\altaffiltext{7}{Alfred P. Sloan Research Fellow}

\slugcomment{DRAFT: \today}

\begin{abstract}

Using a large sample of spectroscopically confirmed Lyman Break
galaxies at $z\sim 3$, we establish for the first time an empirical
relationship between reddening ($\ebmv$), neutral gas covering
fraction ($f_{\rm cov}(\hi)$), and the escape of ionizing (Lyman
continuum, LyC) photons at high redshift.  Our sample includes 933
galaxies at $z\sim 3$, 121 of which have very deep spectroscopic
observations ($\ga 7$\,hrs) in the rest-frame far-UV
($850\la\lambda\la 1300$\,\AA) with the Low Resolution Imaging
Spectrograph on the Keck Telescope.  Based on the high covering
fraction of outflowing optically-thick $\hi$ indicated by the
composite spectra of these galaxies, we conclude that photoelectric
absorption, rather than dust attenuation, dominates the depletion of
LyC photons.  By modeling the composite spectra as the combination of
an unattenuated stellar spectrum including nebular continuum emission
with one that is absorbed by $\hi$ and reddened by a line-of-sight
extinction, we derive an empirical relationship between $\ebmv$ and
$f_{\rm cov}(\hi)$.  Galaxies with redder UV continua have larger
covering fractions of $\hi$ characterized by higher line-of-sight
extinctions.  The absolute escape fraction of Ly$\alpha$ correlates
inversely with $f_{\rm cov}(\hi)$, consistent with the escape of
Ly$\alpha$ through gas- and dust-free lines-of-sight.  Gas covering
fractions based on low-ionization interstellar absorption lines
systematically underpredict those deduced from the $\hi$ lines,
suggesting that much of the outflowing material in these galaxies may
be metal-poor.  We develop a model which connects the ionizing escape
fraction with $\ebmv$, and which may be used to estimate the ionizing
escape fraction for an ensemble of high-redshift galaxies.
Alternatively, direct measurements of the escape fraction for the
composite spectra allow us to constrain the intrinsic
ionizing-to-non-ionizing flux density ratio to be $\langle S(900\,{\rm
  \AA})/S(1500\,{\rm \AA})\rangle_{\rm int} \ga 0.20$, a value that
favors stellar population models that include weaker stellar winds, a
flatter initial mass function, and/or binary evolution.  Lastly, we
demonstrate how the framework discussed here may be used to assess the
pathways by which ionizing radiation escapes from high-redshift
galaxies.

\end{abstract}


\section{INTRODUCTION}
\label{sec:intro}

Significant efforts have focused on determining the sources
responsible for cosmic reionization.  Studies suggesting the apparent
minor role of quasars in keeping the Universe ionized at $z\ga 3$
(e.g., \citealt{madau99, steidel01, fan01, mcdonald01, bolton05,
  bolton07, siana08, glikman11, fontanot12}---but see
\citealt{giallongo15, madau15}) have led the community to assess the
contribution of star-forming galaxies to the ionizing photon budget at
high redshift (e.g., \citealt{yan04, bouwens04, stark07, mclure10,
  bunker10, oesch10, brant10, grazian12, finkelstein12b, ellis13,
  oesch13b, brant15, bouwens15, duncan15}).  These investigations have
advocated broadly for the important role of hitherto undetected
UV-faint galaxies---with their high number densities as inferred from
the steep faint-end slopes of the UV luminosity functions and their
assumed large Lyman continuum (LyC) escape fractions---in dominating
the ionizing background.

However, the direct detection of Lyman continuum radiation, possible
only at modest redshifts ($z\la 3.8$) where the transmissivity of the
inter-galactic medium (IGM) allows such measurements, remains elusive.
Locally, only a handful of objects have been detected with LyC
emission \citep{leitet13, borthakur14, izotov16a, izotov16b,
  leitherer16}.  There are a number of statistical detections in
stacked spectra \citep{steidel01} and individual detections of
candidate LyC emission from both LBGs and Ly$\alpha$-selected samples
\citep{shapley06, iwata09, vanzella10b, nestor11, vanzella12,
  nestor13, mostardi13, debarros16, shapley16} at $z\ga 3$.  However,
foreground contamination has been identified as a significant issue in
interpreting the LyC emission from high-redshift galaxies
\citep{vanzella10a, nestor11, vanzella12, nestor13, siana15,
  mostardi15}.  Nonetheless, at face value, LyC detections at $z\ga
3$, combined with upper limits on the escape fraction locally and at
intermediate redshifts \citep{leitherer95b, hurwitz97, deharveng01,
  siana07, grimes09, cowie09, siana10, bridge10}, have motivated an
interpretation where the LyC escape fraction---i.e., the fraction of
H-ionizing photons that escapes galaxies before being attenuated by
either the IGM or the circum-galactic medium(CGM)---evolves towards
larger values at higher redshift \citep{inoue06, siana07, bridge10,
  siana10}.

Analyses of the limited number of LyC detections, as well as
observations of the Ly$\alpha$ escape fractions in high-redshift
galaxies, suggest that dust reddening and/or gas covering fraction
play an important role in modulating the escape of LyC (and
Ly$\alpha$) photons (e.g., \citealt{shapley03, kornei10, hayes11,
  wofford13, jones13, rudie13, borthakur14, rivera15, trainor15,
  alexandroff15}).  Rest-frame UV spectroscopic studies of
high-redshift galaxies have focused on the propensity of
Ly$\alpha$/LyC escape based on an examination of the gas covering
fraction as typically inferred from saturated low-ionization
interstellar (IS) absorption lines (e.g., \ion{Si}{2}, \ion{O}{1},
\ion{Fe}{2}) that are relatively straightforward to measure (e.g.,
\citealt{shapley03, jones13, trainor15, alexandroff15}).  However,
these transitions may arise in gas of much higher optical
depth---e.g., if the ions are co-spatial with the densest gas---than
that required to significantly attenuate LyC photons ($\log[\nhi/{\rm
    cm}^{-2}]\approx 18$), leading to some ambiguity in the $\hi$
covering fraction.
%
%
Obviously, one should directly measure the column density and covering
fraction of $\hi$ where the Ly$\alpha$/LyC opacity originates.  The
difficulty with this approach, until recently, has been that: (a) the
detection of the $\hi$ absorption lines requires deep far-UV
spectroscopy with blue-sensitive spectrographs, and (b) the proper
interpretation of these lines requires careful corrections for
differential atmospheric refraction relevant at the observed
wavelengths, as well as large samples to average over variations in
Ly$\alpha$ forest blanketing.  Moreover, apart from the anecdotal
evidence of a connection between ionizing escape fraction, {\em
  neutral} gas covering fraction, and reddening, a quantitative study
of how these properties are related to each other has yet to be
undertaken.

To that end, we have used a sample of $z\sim 3$ Lyman Break Galaxies
(LBGs) to place constraints on the two mechanisms that act to suppress
the escape of ionizing photons: photoelectric absorption by neutral
hydrogen ($\hi$) and dust obscuration.  In Reddy et~al. (2016a),
subsequently referred to as Paper~I, we use our sample of $z\sim 3$
LBGs to deduce the shape of the dust attenuation curve to wavelengths
close to the Lyman break.  The resulting curve rises less steeply
towards bluer wavelengths than other commonly adopted attenuation
curves (e.g., \citealt{calzetti00}), implying a factor of $\approx 2$
lower dust attenuation of LyC photons for typical ($L^{\ast}$)
galaxies at $z\sim 3$.  In this paper, we extend these results by
modeling the composite (stacked) UV spectra of $z\sim 3$ galaxies to
assess the neutral gas column densities and covering fractions
relevant for typical star-forming galaxies at these redshifts.  We
then draw a connection between reddening, gas covering fraction, and
ionizing escape fraction.

The outline of this paper is as follows.  In Section~\ref{sec:sample},
we briefly discuss the sample of galaxies and the procedure to
construct their far-UV spectral composites.  The methodology of
fitting models to the composite spectra is presented in
Section~\ref{sec:methodology}.  We discuss the evidence for a
non-unity covering fraction of gas in Section~\ref{sec:nonunity}.  The
modeling of the stacked UV spectra and the correlations found between
continuum reddening and gas covering fraction, column density, and
line-of-sight extinction are considered in
Section~\ref{sec:specfitting}.  We discuss these correlations in the
context of Ly$\alpha$ escape fractions and inferences of gas covering
fractions from low-ionization interstellar absorption lines in
Section~\ref{sec:lines}.  In Section~\ref{sec:escape}, we discuss the
implications of our results in the context of the escape of LyC
photons from high-redshift galaxies, and how the framework established
here can enable powerful constraints on the ionizing stellar
populations of high-redshift galaxies.  All wavelengths are in vacuum.
All magnitudes are expressed in the AB system.  We adopt a cosmology
with $H_{0}=70$\,km\,s$^{-1}$\,Mpc$^{-1}$, $\Omega_{\Lambda}=0.7$, and
$\Omega_{\rm m}=0.3$.

\begin{figure*}
\epsscale{1.19}
\plotone{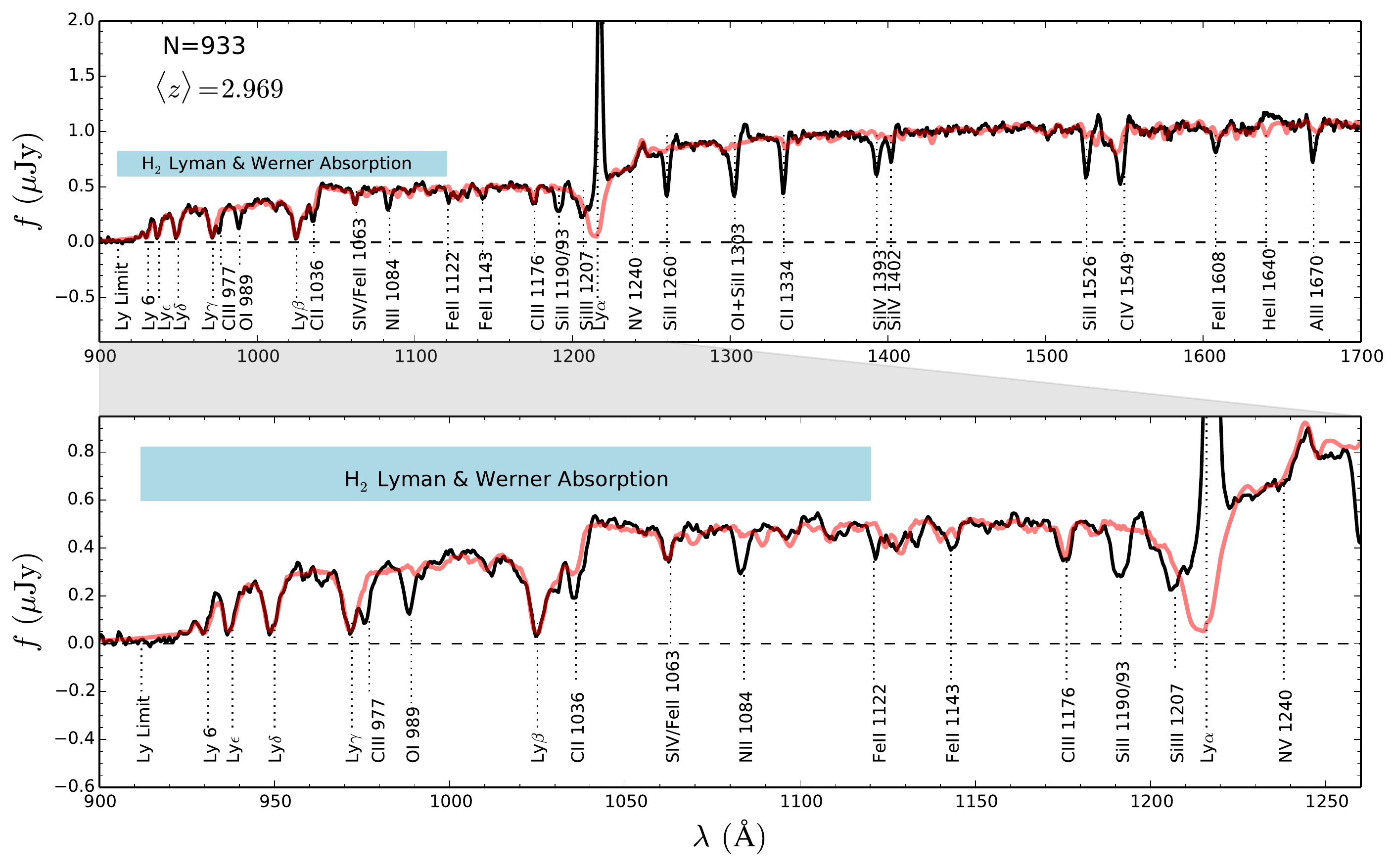}
\caption{{\em Top:} Composite spectrum ({\em black}) of 933 galaxies
  at $z\sim 3$, normalized by the median flux density in the range
  $\lambda=1400-1500$\,\AA.  Several of the most prominent stellar
  photospheric and interstellar absorption lines indicated, along with
  the range of wavelengths ($\lambda=912-1120$\,\AA) corresponding to
  $\molh$ Lyman-Werner absorption.  The best-fit model to the spectrum
  (excluding interstellar metal absorption lines) is shown in red.
  The model assumes an $\hi$/dust covering fraction of $f_{\rm
    cov}(\hi) = 0.96$, a line-of-sight reddening $\ebmv_{\rm
    los}=0.096$ with an SMC extinction curve, $\log[\nhi/{\rm
      cm}^{-2}]=21.0$, an $\molh$ covering fraction of $f_{\rm
    cov}(\molh)=0.03$, and $\log[\nmolh/{\rm cm}^{-2}]=20.9$. {\em
    Bottom:} Same as {\em top} panel, zoomed-in on the region between
  $\lambda=900$ and 1260\,\AA.  The composite shows evidence of
  damping wings in the Lyman series, but line cores that do not reach
  zero intensity as would be expected for the complete covering of
  continuum by optically thick $\hi$.
}
\label{fig:fcov_compall}
\end{figure*}

\section{SAMPLE SELECTION AND REST-FRAME UV SPECTROSCOPY}
\label{sec:sample}

The imaging and spectroscopic survey of $z\sim 3$ LBGs presented in
\citet{steidel03, steidel04} forms the basis for the sample considered
in both Paper~I and the present analysis.  Here, we briefly recount a
few of the salient aspects of our sample, and refer the reader to
Paper~I for additional details.
Candidate $\rs \le 25.5$ galaxies were selected to lie at $z\sim 3$
based on their rest-frame UV colors as measured from ground-based
$U_{\rm n}G\rs$ imaging of 17 fields.  Subsequent shallow spectroscopy
of the candidates ($\sim 1.5$\,hr) with the Low Resolution Imaging
Spectrograph (LRIS; \citealt{oke95}) on the Keck telescope yielded a
sample of 933 star-forming galaxies (AGN were excluded; see Paper~I)
with secure spectroscopic redshifts in the range $2.00\le z_{\rm
  spec}\le 3.75$ (average redshift of $\langle z_{\rm spec}\rangle
=2.969$; see Figure~1 of Paper~I for the histogram of redshifts).  The
$z\sim 3$ UV luminosity function indicates that galaxies in our sample
are typical of those found at these redshifts, with an average
luminosity corresponding to $L^{\ast}$, or an absolute magnitude in
the rest-frame UV ($1700$\,\AA) of $M_{1700} \approx -21$
\citep{steidel99, reddy08, reddy09}.

Of the 933 galaxies, 121 were targeted with very deep ($\simeq
7-10$\,hrs) spectroscopy with the blue arm of LRIS \citep{steidel04},
providing rest-frame far-UV coverage at $850\la \lambda_{\rm rest} \la
1300$\,\AA\, (hereafter referred to as the ``LyC sample'' or the ``LyC
spectra'').  These additional observations were obtained with an
atmospheric dispersion corrector (ADC) to mitigate differential
atmospheric refraction most prominent at blue wavelengths.  The deep
spectroscopy is beneficial not only for the obvious reason of allowing
for greater $S/N$ measurements close to the Lyman Break, but also
enables the more efficient identification of LBGs that are
contaminated by unresolved foreground objects.  The LyC sample
excludes a handful of objects identified as spectroscopic blends (see
Paper~I and Steidel et~al., in prep., for further discussion).  The
rest-frame spectral resolutions of the shallow and deep spectra are
$\approx 1.9$ and $1.5$\,\AA\, respectively.  The individual spectra
were combined to produce a composite spectrum and associated
error spectrum following the procedure presented in Paper~I.  The
signal-to-noise ($S/N$) per resolution element of the final composite
spectrum varies from $S/N \approx 35$ at $\lambda\simeq 920$\,\AA\, to
$\ga 150$ at $\lambda \simeq 1600$\,\AA.
The composite of the 933 galaxies in our sample, including the shallow
and deep spectra, is shown in Figure~\ref{fig:fcov_compall}.

\section{METHODOLOGY}
\label{sec:methodology}

Our analysis relies on the fitting of spectral models to the composite
spectra of $z\sim 3$ galaxies.  The composites enable us to average
over discrete IGM absorption features in order to accurately model the
$\hi$ Lyman absorption lines blueward of Ly$\alpha$.  The methodology
for fitting models to the stacked spectra is similar to that presented
in Paper~I.  Here, we discuss some basic details of the adopted
intrinsic template, the treatment of $\hi$ and $\molh$ absorption, and
the fitting procedures.

\subsection{Intrinsic Template}

Models were fit to the composite spectra in order to deduce the
reddening, average neutral gas column density, and covering fraction
for galaxies in our sample.  To accomplish this, we constructed an
intrinsic template from a combination of the \citet{rix04} models with
the 2010 version of Starburst99 far-UV spectral synthesis models
\citep{leitherer99, leitherer10}.  We adopted a stellar metallicity of
$0.28Z_\odot$, based on current solar abundances \citep{asplund09}, as
this metallicity was found to best reproduce the UV stellar wind
lines, including \ion{C}{3}\,$\lambda$1176 and the
\ion{N}{5}\,$\lambda$1240 and \ion{C}{4}\,$\lambda$1549 P-Cygni
profiles.  Furthermore, a constant star-formation history and an age
of $100$\,Myr were assumed based on results from stellar population
modeling of $z\sim 3$ LBGs presented in \citet{reddy12a}.  Nebular
continuum emission was added to the stellar template to produce the
final ``intrinsic template,'' which we denote by the symbol $m$.

To account for absorption by neutral hydrogen, we added \ion{H}{1}
Lyman series absorption to the intrinsic template assuming a Voigt
profile with a Doppler parameter $b=125$\,km\,s$^{-1}$ (the wings of
the absorption lines are insensitive to the particular choice of $b$)
and varying column densities $\nhi$.\footnote{In
  Section~\ref{sec:newprofile}, we consider a more realistic line
  profile for the $\hi$ derived from the composite spectrum itself.}
The $\hi$ absorption was blueshifted by $300$\,km\,s$^{-1}$ to account
for the observed positions of these lines in the composites.  The
observed blueshift indicates that most of the neutral hydrogen seen in
absorption is outflowing at several hundred km\,s$^{-1}$.  Models that
include $\hi$ absorption are indicated by the symbol $m(\hi)$.
Lastly, we included Lyman-Werner absorption with varying column
densities and covering fractions of $\molh$ as described in Paper~I.
This absorption, which is unresolved, decreases the level of the
continuum at $\lambda \simeq 912 - 1120$\,\AA.  Models that include
both $\hi$ and $\molh$ are referred to by the symbol $m(\hi+\molh)$.

\subsection{Spectral Modeling}

We considered two different cases when fitting the stacked spectra.
In the first case, we calculated the average reddening of the
composite assuming a $100\%$ covering fraction of dust.  This is
analogous to deriving the reddening of a galaxy assuming a starburst
attenuation curve (e.g., \citealt{calzetti00, reddy15}).  In the
subsequent discussion, the ``updated'' versions of the
\citet{calzetti00} and \citet{reddy15} curves refer to those that are
updated for our new measurements of the far-UV shape of the dust curve
as presented in Paper~I.  The intrinsic template was reddened by
various amounts ($\ebmv=0.00-0.60$) assuming the updated attenuation
curves.  The best-fit $\ebmv$ was obtained through $\chi^2$
minimization by comparing the reddened templates---i.e.,
\begin{eqnarray}
m_{\rm final} & = & 10^{-0.4\ebmv k} \times m,
\end{eqnarray}
where $k$ is the attenuation curve---to the composite spectrum in the
``Composite $\ebmv$'' wavelength windows listed in
Table~\ref{tab:waves}.  These windows were chosen to avoid prominent
interstellar and stellar absorption/emission features.  Note that the
windows used to measure $\ebmv$ are unaffected by $\hi$ or $\molh$
absorption, thus obviating the need to fit for this absorption when
determining the reddening of the composite.

\begin{deluxetable}{lc}
\tabletypesize{\footnotesize}
\tablewidth{0pc}
\tablecaption{Wavelength Windows for Fitting}
\tablehead{
\colhead{Type of Fitting} &
\colhead{$\lambda$\,(\AA)}}
\startdata
Composite $\ebmv$\tablenotemark{a} & 1268-1293 \\
 & 1311-1330 \\
 & 1337-1385 \\
 & 1406-1521 \\
 & 1535-1543 \\
 & 1556-1602 \\
 & 1617-1633 \\
\\
$\nhi$\tablenotemark{b} & 967-974 \\
 & 1021-1029 \\
 & 1200-1203 \\
 & 1222-1248 \\
\\
$\nmolh$\tablenotemark{c} & 941-944 \\
 & 956-960 \\
 & 979-986 \\
 & 992-1020
\enddata
\tablenotetext{a}{Windows used to determine the best-fit $\ebmv$
corresponding to a composite spectrum.}
\tablenotetext{b}{Windows used to determine the best-fit column density
$\nhi$ and covering fraction $f_{\rm cov}(\hi)$ of neutral hydrogen.}
\tablenotetext{c}{Windows used to determine the best-fit column density
$\nmolh$ and covering fraction $f_{\rm cov}(\molh)$ of molecular hydrogen.}
\label{tab:waves}
\end{deluxetable}

In Section~\ref{sec:nonunity}, we present evidence from the composite
spectra that indicates a non-unity covering fraction of gas for
galaxies in our sample.  In this second case, we fit models to the
composite spectra assuming non-unity covering fractions of dust and
neutral and molecular hydrogen.  The models were computed as:
\begin{eqnarray}
m_{\rm final} & = & 10^{-0.4\ebmv_{\rm los} k} \times \nonumber \\
& & [(f_{\rm cov}(\hi)-f_{\rm cov}(\molh))m(\hi) + \nonumber \\
& & f_{\rm cov}(\molh)m(\hi+\molh)] + \nonumber \\
& & (1-f_{\rm cov}(\hi))m.
\label{eq:losfit}
\end{eqnarray}
Here, $f_{\rm cov}(\hi)$ is the covering fraction of dust and neutral
gas, and $f_{\rm cov}(\molh)$ is the covering fraction of $\molh$.
This equation assumes that the $\molh$ is shielded by $\hi$ and that
only the portion of the spectrum that emerges through the $\hi$ and
$\molh$ gas is reddened.  As shown in
Sections~\ref{sec:evidencefornonunity} and \ref{sec:specfitting}, this
particular formalism is motivated by the fact that the ISM is expected
to be porous and that the covering fraction of $\molh$ is much smaller
than that of $\hi$ (see Paper I and below).  The last term in
Equation~\ref{eq:losfit} corresponds to the portion of the light that
exits the galaxy unabsorbed, which is simply the intrinsic template
multiplied by $1-f_{\rm cov}(\hi)$.  Finally, $k$ in this equation now
refers to a {\em line-of-sight extinction} curve, in lieu of an
attenuation curve, and $\ebmv_{\rm los}$ refers to the corresponding
line-of-sight reddening.  These adjustments assume that a foreground
uniform dust screen describes the dust-covered portions of the galaxy.

The models were computed assuming a range of line-of-sight reddening,
and covering fractions and column densities of $\hi$ and $\molh$.  In
all cases, the models were smoothed to the spectral resolution of the
composites, and modified by the CGM and IGM opacity appropriate for
the average redshift of each composite.  The average opacity was
determined by randomly drawing line-of-sight absorber column densities
from the distribution measured in the Keck Baryonic Structure Survey
(KBSS; \citealt{rudie13}), and taking an average over many
lines-of-sight.\footnote{Adopting a different prescription for the IGM
  opacity (e.g., \citealt{omeara13, inoue14}) does not affect the
  derived $\hi$ covering fractions, as these fractions are sensitive
  to the ratio of the line depth to the continuum, both of which will
  be affected equally by any changes in the IGM opacity.  However,
  because the $\molh$ covering fraction is calculated by ``matching''
  the model continuum to that of the composite spectrum in the $\molh$
  windows (Table~\ref{tab:waves}), this quantity will suffer a
  systematic bias related to our choice of the IGM+CGM opacity
  prescription (see below).}  The best-fitting $\ebmv_{\rm los}$,
$f_{\rm cov}(\hi)$, and $\nhi$ were found by comparing the models to
the stacked spectra in the ``Composite $\ebmv$'' and ``$\nhi$''
windows listed in Table~\ref{tab:waves}.  As discussed in Paper~I, the
``$\nhi$'' windows were chosen to include the core and the wings of
Ly$\beta$ and Ly$\gamma$, and the wings of Ly$\alpha$, while excluding
regions of metal line absorption (e.g., \ion{C}{3}\,$\lambda 977$ and
\ion{C}{2}\,$\lambda 1036$).  The region between Ly$\alpha$ and
\ion{N}{5}\,$\lambda 1238$ is particularly sensitive to the gas column
density and is included in the fitting windows.  The higher order
Lyman transitions (e.g., Ly$\delta$, Ly$\epsilon$, etc.) were not used
in the fitting due to line overcrowding.  Furthermore, as noted above,
the fitting windows do not include any interstellar metal absorption
lines.
Similarly, the best-fit values of $f_{\rm cov}(\molh)$ and $\nmolh$
were determined by comparing the models to the stacked spectra in
the ``$\nmolh$'' windows listed in Table~\ref{tab:waves}.  As
discussed in Paper~I, the ``$\nmolh$'' windows were chosen to avoid
regions of strong metal line and $\hi$ absorption.

\section{Non-Unity Covering Fraction of Gas in Typical Galaxies at $z\sim 3$}
\label{sec:nonunity}

The apparent porosity of the ISM in galaxies---examples of which have
been seen or suggested in the local Universe (e.g., \citealt{silk97,
  kunth98, oey02, clarke02, bagetakos11, heckman11}), and which may be
enhanced by the high star-formation-rate surface densities
characteristic of high-redshift galaxies (e.g.,
\citealt{razoumov10})---motivates spectral models that allow for less
than a complete covering of the continuum by dust and neutral gas.
Notably, the composite spectrum (Figure~\ref{fig:fcov_compall})
exhibits $\hi$ absorption lines that have damping wings, but which do
not reach zero intensity at the line cores.  The residual flux density
under the $\hi$ lines is $\approx 5\%$ of the median flux density in
the range $\lambda=1400-1500$\,\AA.  For the normalization shown in
Figure~\ref{fig:fcov_compall}, the residual flux density is $\simeq
0.05$\,$\mu$Jy.  As shown in Paper~I, the $\hi$ damping wings are best
reproduced by high column density gas, with $\log[\nhi/{\rm cm}^{-2}]
\approx 21.0$.  At face value, these results suggest a partial
covering fraction of high column density gas.  However, we sought
first to rule out other possible sources for the residual flux
observed at the $\hi$ line cores, namely uncertainties in redshifts
and spectral resolution, and foreground contamination.

\subsection{Redshift and Spectral Resolution Uncertainties}

We determined the extent to which uncertainties in the redshifts and
the resolution of the individual galaxy spectra lead to non-zero
fluxes in the $\hi$ line cores in the stacked spectrum.  To accomplish
this, we added $\hi$ absorption of varying column densities
$\lognhi=20.0-22.0$ to the intrinsic template assuming (as above) a
Voigt profile with a Doppler parameter $b=125$\,km\,s$^{-1}$, and a
$100\%$ covering of gas, resulting in an ``$\hi$-absorbed template''.
Thus, this $\hi$-absorbed template has saturated black $\hi$ line
cores.  The $\hi$-absorbed template was then smoothed to have spectral
resolutions varying from $1.5$\,\AA\, (the average rest-frame
resolution inferred for the LyC spectra) up to a factor of 2 times
worse ($3.0$\,\AA).  For each value of the spectral resolution, we
generated 121 versions of the smoothed $\hi$-absorbed templates---one
for each of the 121 LyC spectra contributing to the composite at
$\lambda < 1150$\,\AA---where each version was shifted in wavelength
according to a normal distribution with $\sigma=125$\,km\,s$^{-1}$.
This velocity spread is based on the uncertainties in the systemic
redshifts when they are derived using interstellar absorption lines
(see \citealt{steidel10}), and the vast majority of the LyC spectra
have systemic redshifts based on the latter.  These 121 versions were
then averaged together to produce a composite model.  Based on these
simulations, we would have had to systematically underestimate the
average resolution of the LyC spectra by $\ga 50\%$ in order to match
the $\hi$ line depths for $\lognhi>20.0$, a highly unlikely situation
given that this $50\%$ systematic difference is much larger than the
rms uncertainty ($<0.1$\,\AA) in determining the spectral resolution
from the widths of the night sky lines.  For the actual value of the
resolution of our spectra and our typical redshift uncertainty, the
cores of the $\hi$ lines remain black, implying that none of the
observed residual flux density can be attributed to spectral
resolution or redshift uncertainties.  We conclude that systematic
uncertainties in spectral resolution cannot account for the residual
line flux density at the $\hi$ line cores.

\subsection{Foreground Contamination}

We ruled out foreground contamination---i.e., low-redshift objects
that are unresolved from the LBGs in the ground-based photometry and
spectroscopy---as a possible origin for the residual line flux
observed at the $\hi$ line cores based on multiple lines of reasoning.
First, we examined the contamination statistics for the sample of LBGs
with spatially-resolved {\em Hubble Space Telescope} ({\em HST})
multi-filter ($UVJH$) and ground-based NB3420 imaging (to target LyC
emission at $z\sim 3$) analyzed in \citet{mostardi13, mostardi15}.
Their sample contains 16 LBGs with {\em HST} imaging, four of which
have NB3420 detections.  Of these four, two were identified as having
foreground contaminants, such that the offending low-redshift
interlopers would not have been resolved separately from the
background LBGs in the ground-based photometry.  Of the 12 objects
that were not detected in the NB3420 imaging, only one suffers from
foreground contamination.  Thus, the contamination rates for NB3420
detections and non-detections are $2/4$ and $1/12$, respectively.
Extrapolating to the full LBG sample of \citet{mostardi13}, where five
of 49 objects were detected in the narrow-band imaging, implies that
the fraction of LBGs with contaminated photometry is $2/4 \times 5/49
+ 1/12 \times 44/49 \approx 0.13$.  Note that what we have just computed is
the foreground contamination {\em rate}.  However, the quantity of
relevance in the present context is not the contamination rate, but
the actual fraction of {\em flux} contributed by foreground
contaminants.

To calculate this flux contamination, we examined the photometry for
the 16 LBGs with {\em HST} imaging from \citet{mostardi15}.  We summed
the $UVJH$ flux densities for the 16 objects, both including and
excluding foreground contaminants that would have been unresolved from
the LBGs in the ground-based imaging.  The ratio of the total flux in
each band including foreground contamination to that excluding this
contamination is $2.38$, $1.02$, $1.03$, and $1.02$, respectively, for
the $UVJH$ bands.  Accounting for the higher foreground contamination
rate among the 16 LBGs ($3/16 \approx 0.19$) relative to the rate
among the full LBG sample ($\approx 0.13$, see above), and adopting
the reasonable assumption that the flux contamination per object is
roughly constant, the aforementioned ratios become $1.92$, $1.01$,
$1.02$, and $1.01$, respectively, for the $UVJH$ bands.  In other
words, the flux contamination fraction for the rest-frame non-ionizing
UV/optical flux is $<2\%$.  The flux contamination fraction in the
$U$-band is substantially larger, a point that we revisit in
Section~\ref{sec:escape}.  Based on these statistics, the
contaminating flux density across the wavelength range spanning the
Ly$\beta$, Ly$\gamma$, and Ly$\delta$ lines is a factor of $\approx 5$
lower than the residual flux densities measured at the aforementioned
line centers.

Note that the previously discussed contamination rates were calculated
for LBGs that were spectroscopically confirmed using relatively
shallow spectroscopic data \citep{reddy08, steidel11, mostardi13}.
Spectroscopic blends may not be as readily identifiable in these
shallow spectra compared to the substantially deeper LyC spectra
analyzed here.  As indicated in Section~\ref{sec:sample}, the $\simeq
7-10$\,hrs spectroscopic integrations obtained with LRIS-B allowed us
to remove a handful ($\approx 5\%$) of LBGs that were confused with
foreground interlopers.  In fact, this contaminating rate is
consistent with the expected number of chance superpositions of
foreground objects, as discussed in Steidel et~al. (in prep), implying
that the present sample has been expunged of most contaminants.  
Thus, the contamination rates calculated above should be regarded as
conservative upper limits.

There is additional evidence that suggests that the residual line flux
at the $\hi$ line centers is intrinsic to the LBGs.  Namely, in the
subsequent analysis, we show that this residual flux varies
systematically with $\ebmv$ such that redder galaxies exhibit lower
levels of residual emission (Section~\ref{sec:specfitting}), a trend
that would not be expected if foreground interlopers were dominating
the observed flux.  For example, of the three LBGs identified with
foreground contamination from the sample of \citet{mostardi15}, two
have interlopers that result in slightly {\em redder} colors compared
to the corrected photometry, and one has an interloper that results in
a color that is indistinguishable from the corrected color.
Notwithstanding the small statistics, it follows that an increase in
contamination fraction for galaxies with redder $\ebmv$ should lead to
a corresponding {\em increase} in residual flux, a trend that is
opposite to the one actually observed (Section~\ref{sec:specfitting}).

\subsection{Summary of the Evidence for a Non-Unity Covering Fraction of $\hi$}
\label{sec:evidencefornonunity}

To summarize our arguments, we find that systematic uncertainties in
the resolution of our spectra, combined with random uncertainties in
systemic redshifts, cannot explain the level of residual flux seen in
the stacked spectra.  Furthermore, calculations of the flux
contamination fraction for the sample of LBGs examined in
\citet{mostardi13, mostardi15} imply that $>80\%$ of the residual flux
must be intrinsic to the LBGs and not a result of foreground
interlopers.  If the statistics of how the colors of LBGs are altered
due to foreground interlopers extend to larger samples, then
significant contamination would be expected to lead to a higher
residual flux for redder galaxies, contrary to the trend that we
observe, as discussed in Section~\ref{sec:specfitting}.

We conclude with some words of caution regarding the covering
fractions derived here.  First, while the composite may indicate
$f_{\rm cov}(\hi)<1$, low resolution spectroscopy may mask the
presence of very narrow absorption components that are opaque to LyC
radiation.\footnote{These narrow absorption components cannot have
  column densities much larger than the ones deduced from the
  composite spectrum, otherwise their presence would be revealed by
  substantially broader Lorentzian wings.}  However, if the presence
of such narrow (and black) components were the rule, rather than the
exception, for galaxies in our sample, we would expect little
correlation between the covering fraction derived from low resolution
spectra and the Ly$\alpha$ escape fraction, counter to our
observations (Sections~\ref{sec:specfitting} and \ref{sec:lines}).

\begin{figure*}
\plotone{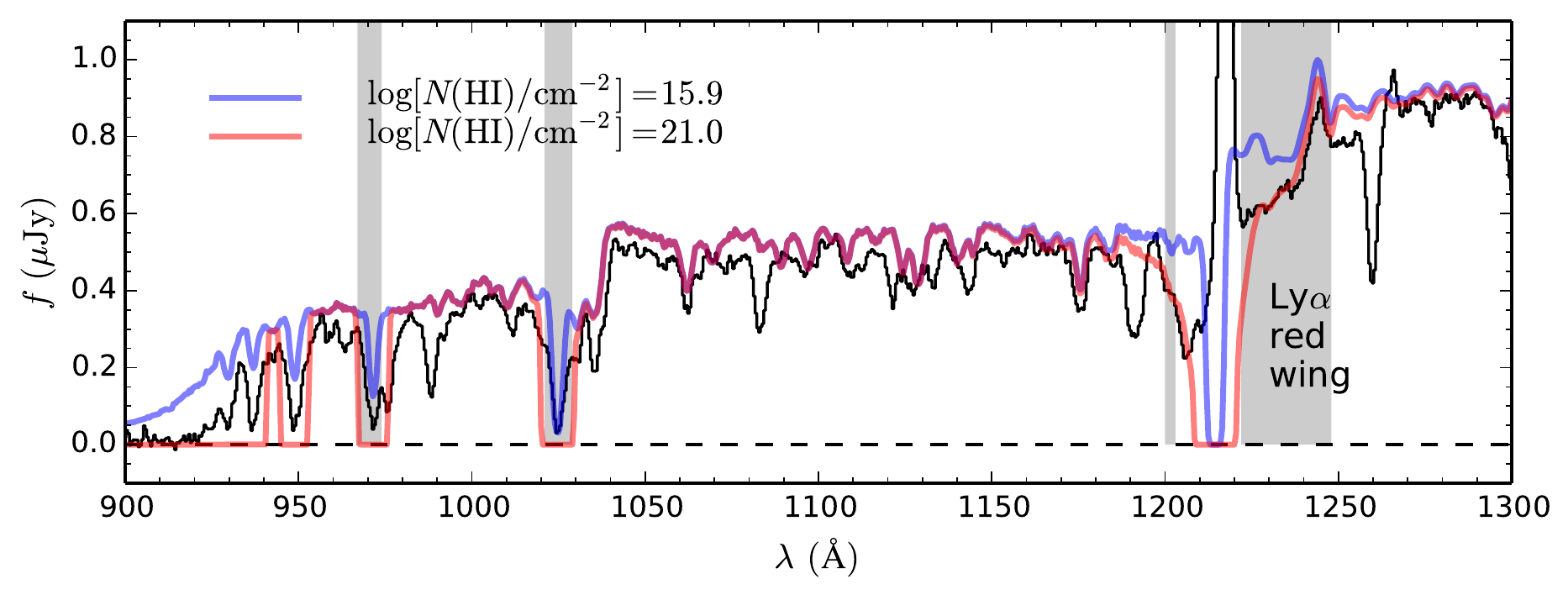}
\caption{Model fits to the $\ebmv$ and $\hi$ lines---assuming $100\%$
  covering of dust and neutral gas with two different column
  densities, and assuming that the $\hi$ lines are described by a
  Voigt profile---to the composite spectrum of all galaxies in our
  sample (Figure~\ref{fig:fcov_compall}).  The models have been
  smoothed to the same spectral resolution as the composite spectra.
  For reference, the 4 windows used to fit the Lyman series lines are
  indicated by the shaded regions (Table~\ref{tab:waves}).  The
  windows around Ly$\alpha$ are designed to exclude the regions with
  strong Ly$\alpha$ emission, \ion{Si}{3} $\lambda 1207$ absorption,
  and include the red wing of Ly$\alpha$.  A low column density of
  $\lognhi=15.9$ matches the depth of the Ly$\beta$ line, but fails to
  reproduce the red damping wing of Ly$\alpha$ (which is blended with
  the \ion{N}{5} P-Cygni absorption) and the depths of the higher
  Lyman series lines.  A high column density of $\lognhi=21.0$ is able
  to reproduce the red damping wing of Ly$\alpha$, but implies
  saturated $\hi$ lines.  The roughly equal depths of the Lyman series
  lines with residual intensity at the line cores implies a partial
  covering fraction of high column density neutral gas.}
\label{fig:covfracexample}
\end{figure*}

\begin{figure}
\epsscale{1.00}
\plotone{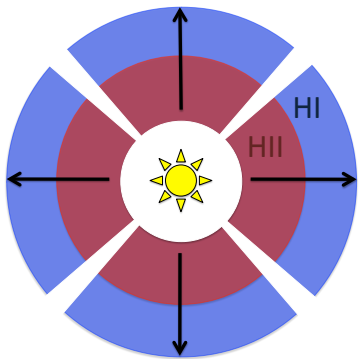}
\caption{Cartoon configuration of a galaxy with a porous ISM and hence
  non-unity covering fraction of dust and gas.  In this geometry, the
  ionized region is bounded by neutral gas (i.e., corresponding to a
  radiation-bounded nebula), and both the ionized and neutral
  components are outflowing, as indicated by the arrows.  Note that
  the composite spectrum (Figure~\ref{fig:fcov_compall}) also
  indicates non-negligible $\hi$ opacity at systemic redshift.  Holes
  in the ISM, such as those formed by supernovae, result in regions of
  very low gas and dust column density.  In this framework, the dust
  is coincident with the gas, but may be inhomogeneously distributed
  between the neutral and ionized phases.}
\label{fig:cartoon}
\end{figure}

\begin{figure*}
\epsscale{1.10}
\plotone{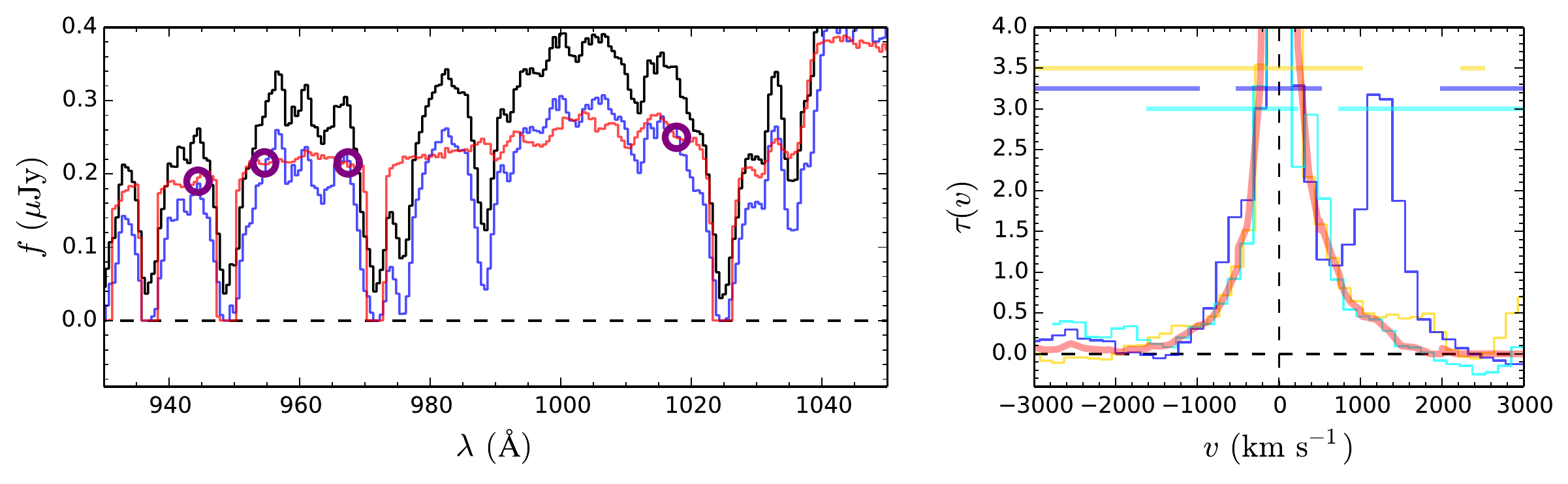}
\caption{{\em Left:} Composite spectrum of all galaxies in our sample
  ({\em black}), compared with the result when subtracting $(1-f_{\rm
    cov}(\hi))\times m$, where $m$ is the intrinsic template, from the
  composite spectrum ({\em blue}) and the model fit to the composite
  spectrum ({\em red}).  The ``continuum points'' used to normalize
  the Ly$\beta$, Ly$\gamma$, and Ly$\delta$ line profiles are
  indicated by the open circles, and were determined by the
  intersection of the {\em blue} and {\em red} curves.  {\em Right:}
  Optical depths of the Ly$\beta$ ({\em orange}), Ly$\gamma$ ({\em
    blue}), and Ly$\delta$ ({\em cyan}) lines as a function of
  velocity.  The windows over which the average optical depth profile
  ({\em thick orange line}) was computed are indicated by the
  horizontal {\em yellow}, {\em blue}, and {\em cyan} lines for the
  Ly$\beta$, Ly$\gamma$, and Ly$\delta$ lines, respectively. }
\label{fig:subtract}
\end{figure*}

Second, the method for computing $f_{\rm cov}(\hi)$ assumes that the
different velocity components of the $\hi$ gas are spatially
co-incident when viewed in projection.  If this is not the
case---i.e., if optically-thick gas that absorbs far from line center
covers sightlines that are disjoint from those where the absorption at
line center dominates---then $f_{\rm cov}(\hi)$ deduced based on the
overall depth of the $\hi$ absorption will underestimate the true
$\hi$ covering fraction.  As we show in Section~\ref{sec:escape}, the
two effects just mentioned (i.e., unresolved narrow absorption and
spatially disjoint gas at different velocities) must be limited in
scope.  Even so, in light of this discussion, one must consider that
the covering fractions derived in our subsequent analysis may be lower
limits to the true fractions.

To demonstrate the requirement of a non-unity covering fraction, we
modeled the composite spectrum of all galaxies in our sample
(Figure~\ref{fig:fcov_compall}) using the same procedure as above in
fitting $\ebmv$ and $\hi$ (we ignored $\molh$ for this exercise), but
where we now assumed a unity covering fraction of both dust and gas.
The depth of the Ly$\beta$ line can be reproduced with $\lognhi=15.9$,
but such a low column density does not match the combined red damping
wing of Ly$\alpha$ and P-Cygni \ion{N}{5} absorption, nor the depths
of the higher Lyman series lines (Figure~\ref{fig:covfracexample}).
Alternatively, the assumption of $\lognhi=21.0$ gas reproduces the
combined red damping wing of Ly$\alpha$ and P-Cygni \ion{N}{5}
absorption, but implies optically-thick gas and hence saturated $\hi$
lines that should reach zero intensity at the line cores for a unity
covering fraction.

Thus, it is clear that neither low or high column density gas with
complete covering can simultaneously reproduce the residual flux at
the $\hi$ line centers and the damping wings of Ly$\alpha$.  Rather,
the roughly equal depths of the Lyman series lines combined with the
damping wings of Ly$\alpha$ suggest optically thick gas with
$\lognhi\ga 21.0$ and $f_{\rm cov}(\hi)<1$.
%
Consequently, the depths of the $\hi$ lines and the red damping wing
of Ly$\alpha$ are most sensitive to changes in the covering fraction
of optically thick $\hi$.

\subsection{A Physical Picture for a Non-Unity Covering
Fraction of Gas and Dust}

From a physical standpoint, a partial covering of dust and gas can
result from supernovae chimneys or galactic fountains, whereby
supernovae explosions heat the ISM and drive hot gas into the galactic
halo (e.g., \citealt{putman12} and references therein).  Aside from
direct observations of such hot halos in the Milky Way and other
nearby galaxies (e.g., \citealt{lynds63, shapiro76, lehnert95,
  armus95, strickland04b}), the higher star-formation-rate surface
densities of, and the ubiquitous presence of outflows in,
high-redshift galaxies may increase the porosity of the ISM.
Furthermore, simulations of high-redshift galaxies show that the LyC
escape fraction is essentially independent of dust obscuration,
with the ionizing photons escaping the galaxies through channels
that are essentially free of dust \citep{gnedin08, razoumov10, ma16},
as depicted in Figure~\ref{fig:cartoon}.  In this cartoon picture, the
column of gas is sufficient to completely bound the ionized region by
neutral gas, where the bulk of the neutral and ionized gas along the
line-of-sight are entrained in an outflow (see also Figure~1 in
\citealt{zackrisson13}).  

\section{SPECTRAL FITTING RESULTS}
\label{sec:specfitting}

\subsection{An Initial Fit to the Full Composite Spectrum}


Using the procedure laid out in Section~\ref{sec:methodology}, we fit a
spectral model to the composite of all galaxies (full composite), assuming
a non-unity covering fraction of dust and gas, and an SMC extinction curve.
The model was obtained by 
%
simultaneously varying $\ebmv_{\rm los/SMC}$, $f_{\rm cov}(\hi)$,
$\nhi$, $f_{\rm cov}(\molh)$, and $\nmolh$, until a best-fit was
reached.  The best-fit parameters are $\ebmv_{\rm los/SMC}=0.096$,
$\log[\nhi/{\rm cm}^{-2}]=21.0$, $f_{\rm cov}(\hi) = 0.96$,
$\log[\nmolh/{\rm cm}^{-2}]=20.9$, and $f_{\rm cov}(\molh)=0.03$.  By
perturbing the composite spectrum by its errors and refitting these
realizations, we estimated the errors in $\ebmv_{\rm los}$, $f_{\rm
  cov}(\hi)$, and $\log[\nhi/{\rm cm}^{-2}]$ to be $\sigma(\ebmv_{\rm
  los}) \approx 0.01$, $\sigma(f_{\rm cov}(\hi)) \approx 0.01$, and
$\sigma(\log[\nhi/{\rm cm}^{-2}]) \approx 0.20$\,dex, respectively.
Nominally, these errors include the covariance between the parameters.
However, as $f_{\rm cov}(\hi)$ is determined primarily by the depth of
the $\hi$ lines, and $\nhi$ is constrained primarily by the wings,
there is negligible covariance between the covering fraction and
column density of $\hi$.  Furthermore, as noted in Paper~I, the
$\molh$ lines are unresolved and therefore we cannot accurately
determine the errors in the $\molh$ covering fraction and column
density.  In other words, for the models to match the continuum level
of the stacked spectrum in the ``$\nmolh$'' windows, the $\molh$
column density may be increased while the covering fraction is
decreased, or vice versa.  As noted above, $f_{\rm cov}(\molh)$ may
also suffer a systematic bias depending on which IGM opacity
prescription we adopt.

\begin{figure*}
\epsscale{1.10}
\plotone{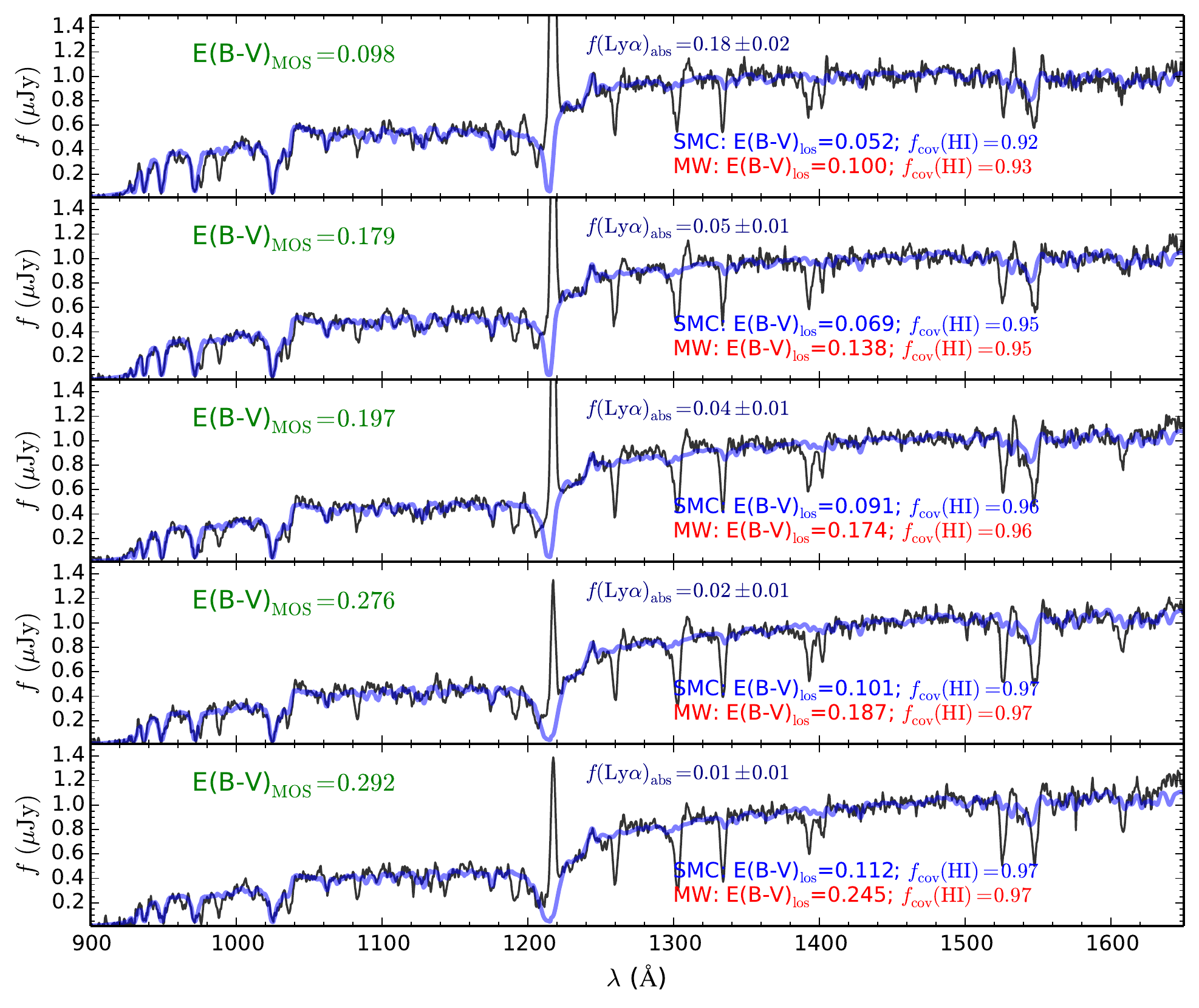}
\caption{Best-fitting models assuming a non-unity covering fraction of
  neutral gas and dust, and the SMC extinction curve ({\em blue}), to
  the composite spectra ({\em black}), in five bins of $\ebmv$.  The
  fits assuming the MW extinction curve have been omitted for clarity,
  but they are similar to those obtained with the SMC extinction
  curve. The best-fitting $\ebmv$ assuming the updated \citet{reddy15}
  attenuation curve and a unity covering fraction of dust are
  indicated in the upper left-hand corner of each panel.  Also
  indicated are the best-fit $\ebmv_{\rm los}$ and $f_{\rm cov}(\hi)$,
  assuming the SMC and Milky Way extinction curves.
  Uncertainties on the parameters are listed in
  Table~\ref{tab:losfit}, and were estimated by perturbing the
  composite spectra according to their error spectra and re-fitting
  the models (see text).  Each panel indicates the corresponding
  absolute escape fraction of Ly$\alpha$ flux.}
\label{fig:confint}
\end{figure*}

\begin{deluxetable*}{lcccccccc}
\tabletypesize{\footnotesize}
\tablewidth{0pc}
\tablecaption{Best-Fit $\ebmv$ and $f_{\rm cov}(\hi)$}
\tablehead{
\colhead{$\ebmv_{\rm MOS}$\tablenotemark{a}} & 
\colhead{$\ebmv_{\rm Calz}$\tablenotemark{a}} &
\colhead{$\ebmv_{\rm los/SMC}$\tablenotemark{b}} &
\colhead{$f_{\rm cov}(\hi)$\tablenotemark{b}} &
\colhead{$\lognhi$\tablenotemark{b}} &
\colhead{} &
\colhead{$\ebmv_{\rm los/MW}$\tablenotemark{c}} &
\colhead{$f_{\rm cov}(\hi)$\tablenotemark{c}} & 
\colhead{$\lognhi$\tablenotemark{c}}}
\startdata
$0.098\pm 0.005$ & $0.104\pm 0.006$ & $0.052\pm 0.002$ & $0.92$ & $20.3$ & & $0.100\pm 0.002$ & $0.93$ & $20.4$ \\
$0.179\pm 0.005$ & $0.184\pm 0.006$ & $0.069\pm 0.002$ & $0.95$ & $20.3$ & & $0.138\pm 0.002$ & $0.95$ & $20.5$ \\
$0.197\pm 0.005$ & $0.203\pm 0.008$ & $0.091\pm 0.001$ & $0.96$ & $20.5$ & & $0.174\pm 0.002$ & $0.96$ & $20.8$ \\ 
$0.276\pm 0.006$ & $0.285\pm 0.005$ & $0.101\pm 0.001$ & $0.97$ & $21.0$ & & $0.187\pm 0.004$ & $0.97$ & $21.2$ \\
$0.292\pm 0.006$ & $0.301\pm 0.007$ & $0.112\pm 0.001$ & $0.97$ & $21.0$ & & $0.245\pm 0.004$ & $0.97$ & $21.4$ 
\enddata
\tablenotetext{a}{Continuum reddening assuming the updated \citet{reddy15} and \citet{calzetti00} attenuation curves.}
\tablenotetext{b}{Best-fitting $\ebmv_{\rm los}$, $f_{\rm cov}(\hi)$, and $\lognhi$, assuming a non-unity covering fraction of neutral gas and dust and the SMC extinction curve.
The uncertainties in the $\hi$ covering fractions are $\sigma(f_{\rm cov}(\hi)) = 0.01$.  The uncertainties in $\lognhi$ are $\sigma(\lognhi)=0.2$\,dex.}
\tablenotetext{c}{Best-fitting $\ebmv_{\rm los}$, $f_{\rm cov}(\hi)$, and $\lognhi$, assuming a non-unity covering fraction of neutral gas and dust and the Milky Way extinction curve.
The uncertainties in the $\hi$ covering fractions are $\sigma(f_{\rm cov}(\hi)) = 0.01$.  The uncertainties in $\lognhi$ are $\sigma(\lognhi)=0.2$\,dex.}
\label{tab:losfit}
\end{deluxetable*}

\subsection{Derivation of the $\hi$ Line Profile}
\label{sec:newprofile}

As shown in Figure~\ref{fig:covfracexample} and in Paper~I, a Voigt
profile provides a poor description for the shape of the $\hi$ lines
(i.e., the Voigt profile underestimates the width of the lines when
simultaneously matching their depths).  To obtain a better fit for
these lines, we derived the shape of the $\hi$ line profile from the
composite spectrum itself.
To recover the $\hi$ line profile, the intrinsic spectrum was
multiplied by $1-f_{\rm cov}(\hi)$, modified for the opacity of the
IGM, and subtracted from both the composite and its model fit, with
the resulting spectra shown in Figure~\ref{fig:subtract}.  This step
allows us to remove the contribution of unabsorbed continuum prior to
modeling the $\hi$ lines.  We then used the model fit subtracted by
the intrinsic spectrum (i.e., red curve in Figure~\ref{fig:subtract})
to define ``continuum'' points by which the Ly$\beta$, Ly$\gamma$, and
Ly$\delta$ line profiles were normalized.  There is a single continuum
point each for Ly$\beta$ and Ly$\gamma$ and we treated the continuum
as constant across these line profiles.  For Ly$\delta$, the continuum
was estimated as a linear function passing through the two continuum
points bracketing the line.  The normalized line profiles, transformed
into velocity space (where the minimum of each absorption trough is
forced to zero velocity), and cast in terms of the optical depth, are
shown in Figure~\ref{fig:subtract}.

The optical depth profile shown in Figure~\ref{fig:subtract} was
assumed for all of the Lyman series lines, except Ly$\alpha$.  Because
the damping wings are weaker and more difficult to measure for the
higher Lyman series lines due to line crowding, we modeled the damping
wings of Ly$\alpha$ using a range of $\hi$ column densities with a
Doppler parameter $b=125$\,km\,s$^{-1}$---as noted earlier, the wings
of the absorption are insensitive to the particular choice of $b$.

\subsection{Composite Fitting}

In order to investigate trends between continuum color excess and
covering fraction, we divided the galaxies in our sample into five
bins of $\ebmv$.  Specifically, the intrinsic template was reddened
with different values of $\ebmv$ assuming the \citet{reddy15}
attenuation curve and a $100\%$ covering fraction of dust, shifted to
the observed frame for the redshift of each galaxy, multiplied by the
IGM opacity relevant for that redshift, and then convolved with the
$G$ and $\rs$ filters.  The best-fit $\ebmv$ was taken to be the value
that resulted in a model-generated $\gmr$ that best matched the
observed $\gmr$ color, where the latter was corrected for Ly$\alpha$
emission or absorption as measured from the spectra.  Composite
spectra for objects in each bin were constructed according to the
procedures laid out in Paper~I and Section~\ref{sec:sample}.

In fitting the stacked spectra assuming a porous ISM, we considered
both the SMC and Milky Way extinction curves in the modeling (i.e.,
for the {\em line-of-sight} reddening), to span the possible shapes of
the extinction curve in both low and high metallicity environments.
The fitting was accomplished by simultaneously varying $\ebmv_{\rm
  los}$, $\nhi$ (which affects the degree of damping of the Ly$\alpha$
line), $f_{\rm cov}(\hi)$ (which affects the depths of all the Lyman
series lines), and $f_{\rm cov}(\molh)$, until the differences between
the composite spectra and the models were minimized in the wavelength
regions indicated by the ``Composite $\ebmv$'', ``$\nhi$'', and
``$\nmolh$'' entries in Table~\ref{tab:waves}.  As the $\molh$ lines
are unresolved, we fixed $\lognmolh=20.9$ in the fitting.  We verified
that the covering fractions of neutral gas obtained with and without
including $\molh$ in the model fits were similar within $\approx 1\%$.
The covering fractions of $\molh$ are significantly smaller, with a
typical value of $f_{\rm cov}(\molh) \la 0.05$.  Additionally, each
composite spectrum was fit assuming a $100\%$ covering fraction of
dust in order to derive its $\ebmv$.  The $\ebmv$ computed from the
composite spectra agree well with the average $\ebmv$ of objects
contributing to those spectra.  For simplicity in the following
discussion, the color excess derived assuming a unity covering
fraction of dust is referred to as the ``continuum reddening'',
$\ebmv$, while the reddening deduced assuming a partial covering
fraction of dust is referred to as the ``line-of-sight reddening'',
$\ebmv_{\rm los}$.  The continuum reddenings derived assuming the
updated attenuation curves of \citet{reddy15} and \citet{calzetti00}
are denoted by ``$\ebmv_{\rm MOS}$'' and ``$\ebmv_{\rm Calz}$,''
respectively.

The fits assuming the SMC extinction curve and a non-unity covering
fraction of dust and neutral gas are shown in
Figure~\ref{fig:confint}, and the best-fit parameters and their errors
are summarized in Table~\ref{tab:losfit}.  The errors were calculated
as the $1$\,$\sigma$ dispersion in the values obtained by perturbing
the composite spectra according to their error spectra and re-fitting
these realizations.  Not surprisingly, the new $\hi$ optical depth
profile adopted for Ly$\beta$ and the higher Lyman series transitions
(Section~\ref{sec:newprofile}) provides a significantly better fit for
these lines---as well as the continuum level at $\lambda <
950$\,\AA---than a single Voigt profile, though the inferred covering
fractions remain identical (see Figure~\ref{fig:fcov_compall} for the
model fit to the full composite).

\begin{figure*}
\epsscale{1.10}
\plotone{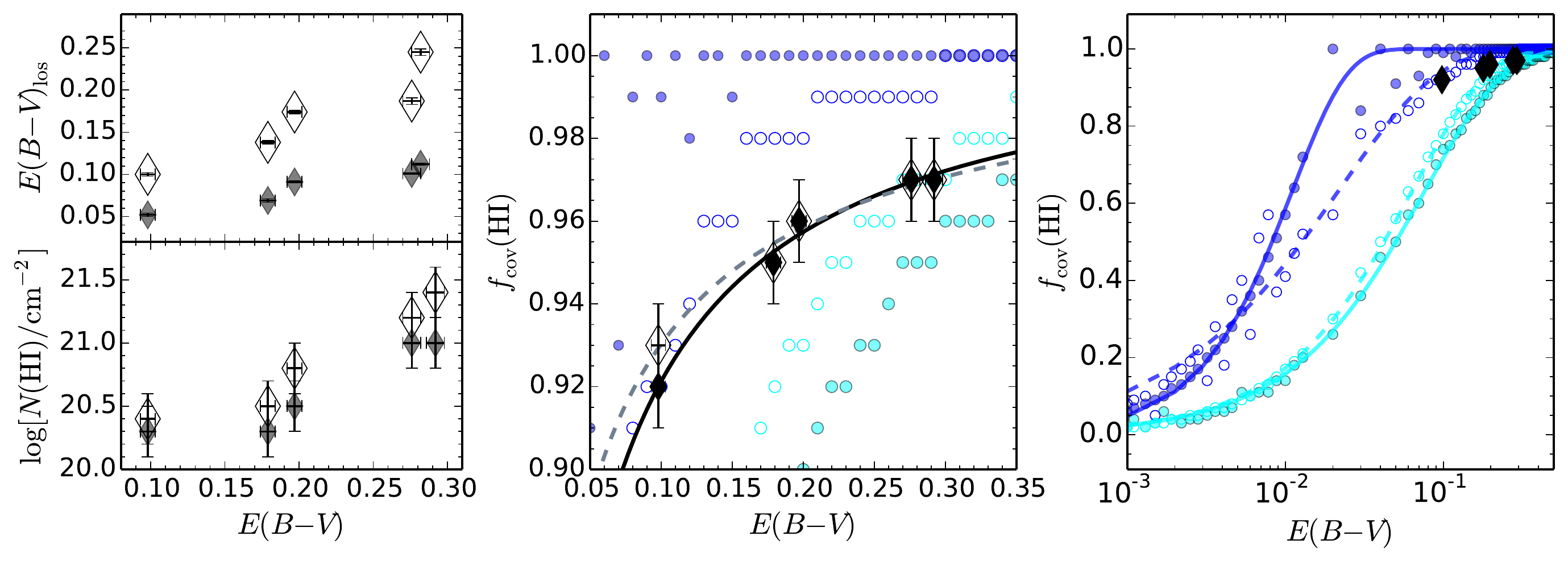}
\caption{{\em Left:} Relation between continuum and line-of-sight
  reddening ({\em top}), and continuum reddening and column density of
  neutral hydrogen ({\em bottom}), assuming a non-unity covering
  fraction of dust and the SMC ({\em filled diamonds}) and Milky Way
  ({\em open diamonds}) extinction curves.  {\em Middle:} Relation
  between $\hi$ covering fraction and $\ebmv$.  The {\em filled} and
  {\em open diamonds} show the measurements obtained directly from the
  composite spectra assuming the SMC and Milky Way extinction curves,
  respectively.  The {\em solid} and {\em dashed lines} indicate the
  best fits to these measurements.  The circles denote model
  expectations of the relationship between covering fraction and
  $\ebmv$.  The {\em cyan} and {\em blue circles} denote the results
  when assuming the updated and original \citet{reddy15} attenuation
  curves, respectively, for computing the continuum reddening.  The
  {\em open} and {\em filled circles} correspond to the results when
  we assume the SMC and MW extinction curves, respectively, for
  computing the covering fractions.  {\em Right:} Same as the middle
  panel, but showing a greater dynamic range in both continuum
  reddening and covering fraction.  Also indicated are the best-fit
  relations ({\em solid} and {\em dashed} lines) to the model points
  assuming the functional form expressed in
  Equation~\ref{eq:functional}.}
\label{fig:covfrac}
\end{figure*}

\begin{deluxetable*}{llcccc}
\tabletypesize{\footnotesize}
\tablewidth{0pc}
\tablecaption{Summary of Correlation Analysis and Fits\tablenotemark{i}}
\tablehead{
\colhead{$x$ variable} &
\colhead{$y$ variable} &
\colhead{$a$} & 
\colhead{$b$} &
\colhead{$\rho$\tablenotemark{ii}} &
\colhead{$p$\tablenotemark{iii}}}
\startdata
$\ebmv$ & $f_{\rm cov}(\hi)$ & $-5.214\pm0.736$\tablenotemark{iv} & $0.312\pm0.074$\tablenotemark{iv} & $0.90$ & $0.04$ \\
$\log\ebmv$ & $\lognhi$ & $1.60\pm0.53$\tablenotemark{v} & $21.76\pm0.39$\tablenotemark{v} & $0.85$ & $0.07$ \\
$\log\ebmv$ & $\log\ebmv_{\rm los}$ & $0.614\pm0.025$\tablenotemark{vi} & $-0.632\pm0.015$\tablenotemark{vi} & $0.96$ & $0.01$
\enddata
\tablenotetext{i}{Parameters are presented for the correlations obtained when
assuming the updated \citet{reddy15} stellar attenuation curve and the SMC extinction
curve.  Other combinations of the attenuation/extinction curves result in correlations with
similar fits, with similar Pearson correlation coefficients and similar probabilities for
null correlations between the considered quantities.}
\tablenotetext{ii}{Pearson correlation coefficient.}
\tablenotetext{iii}{Probability of a null correlation.}
\tablenotetext{iv}{Free parameters $a$ and $b$, as in Equation~\ref{eq:functional}.}
\tablenotetext{v}{Slope and intercept, denoted by $a$ and $b$, respectively, in the linear
fit to $\lognhi$ vs. $\log\ebmv$.}
\tablenotetext{vi}{Slope and intercept, denoted by $a$ and $b$, respectively, in the linear
fit to $\log\ebmv_{\rm los}$ vs. $\log\ebmv$.}
\label{tab:covfracfits}
\end{deluxetable*}

\subsection{Summary of the Spectral Fitting Results}
\label{sec:correlations}

Several notable results from the fitting are summarized in
Figure~\ref{fig:covfrac} and Table~\ref{tab:covfracfits}.  Significant
correlations are found between the continuum reddening and the
line-of-sight reddening, column density of $\hi$, and the covering
fraction of $\hi$.  For simplicity in the subsequent analysis, and
unless indicated otherwise, we assume the updated \citet{reddy15}
curve for the continuum reddening, and the SMC extinction curve for
deriving the line-of-sight extinction and the neutral gas covering
fraction and column density.  Other combinations of the
extinction/attenuation curves yield similar results to those obtained
with the default choices specified above.

To compute the significance of the correlations between $\ebmv$ and
$\ebmv_{\rm los}$, $\nhi$, and $f_{\rm cov}(\hi)$, we calculated the
Pearson correlation coefficient and probability of a null correlation
between the various quantities while perturbing the measured values
many times according to their measurement uncertainties.  The values
reported in Table~\ref{tab:covfracfits} are the median Pearson
correlation coefficients and median probabilities of a null
correlation between $\ebmv$ and $f_{\rm cov}(\hi)$, $\nhi$, and
$\ebmv_{\rm los}$.  The inferred $\hi$ covering fractions assuming a
partial covering fraction of both dust and gas are $\ga 92\%$,
substantially larger than those deduced assuming a unity covering
fraction of dust ($f_{\rm cov}(\hi) \simeq 0.70-0.80$; see Paper~I).
In the latter case, as the entire spectrum is reddened including the
uncovered portion, the line depths can be reproduced with smaller
covering fractions of $\hi$.  With a non-unity covering fraction of
dust, larger $\hi$ covering fractions are required to reproduce the
depths when unreddened continuum is contributed from the uncovered
portion of the galaxy spectrum.  At any rate, our results imply that
the reddening of the UV continuum slope reflects an increasing
covering fraction of $\hi$ and dust, where the covering gas is also
characterized by a higher line-of-sight reddening.

\begin{figure}
\epsscale{1.00}
\plotone{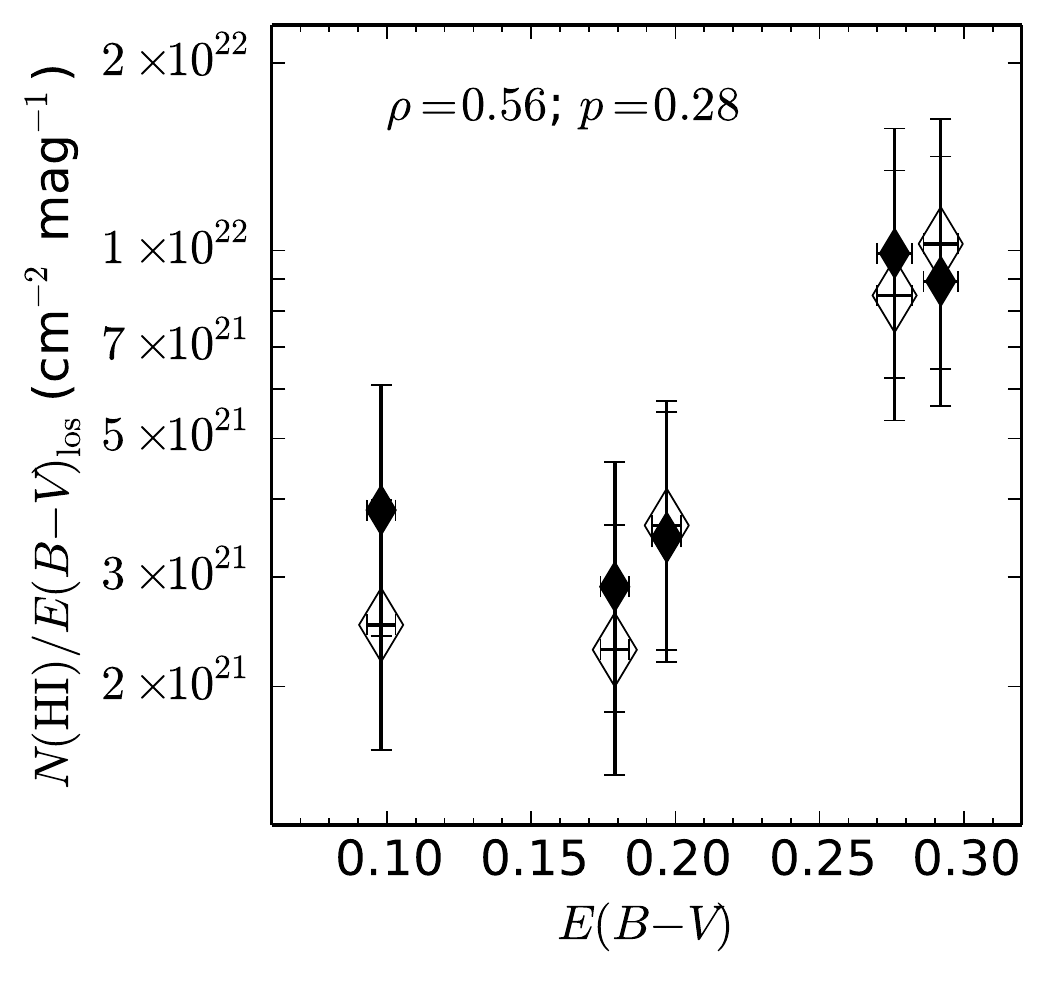}
\caption{Ratio of the neutral gas column density and line-of-sight
  reddening, assuming the SMC ({\em filled diamonds}) and Milky Way
  ({\em open diamonds}) extinction curves, as a function of continuum
  reddening.  The Pearson correlation coefficient and probability of a
  null correlation between $\nhi/\ebmv_{\rm los}$ and continuum
  reddening are indicated in the panel for the SMC case, and are
  similar to those obtained for a Milky Way extinction curve.}
\label{fig:nhiebmv}
\end{figure}

Furthermore, the column density of $\hi$ increases with continuum
reddening.  The ratio of the neutral gas column density and the
line-of-sight reddening is only marginally correlated with continuum
reddening with a Pearson correlation coefficient of $\rho = 0.56$ and
a probability of null correlation of $p=0.28$
(Figure~\ref{fig:nhiebmv}).  The mean ratio of the neutral gas column
density and the line-of-sight reddening for our sample is
$\langle\nhi/\ebmv_{\rm los}\rangle=(5.8\pm3.3)\times 10^{21}$ and
$(5.4\pm3.7)\times 10^{21}$\,cm$^{-2}$\,mag$^{-1}$ for the SMC and
Milky Way extinction curves, respectively, where the errors represent
the standard deviation in the five values contributing to the mean.
For context, the average values found for SMC and MW sightlines are
$\sim 2.1\times 10^{22}$ and $\sim 4.5\times
10^{21}$\,cm$^{-2}$\,mag$^{-1}$, respectively \citep{welty12}.  A note
of caution is that the $\langle\nhi/\ebmv_{\rm los}\rangle$ values
calculated here are analogous to, but not the same as, the gas-to-dust
ratio.  Specifically, we do not make any accounting of either the
molecular or ionized gas in our column density estimates, and much of
the line-of-sight reddening can be attributed to dust in the
outflowing {\em ionized} gas in high-redshift galaxies (see
Section~\ref{sec:isabslines}).


\section{Inferences from Ly$\alpha$ and the Low- and High-Ionization
Interstellar Absorption Lines}
\label{sec:lines}

The discussion in the previous section focused on the parameters
obtained from fitting the composite spectra, including the continuum
and line-of-sight reddening, and the covering fraction and column
density of $\hi$.  We now turn our attention to other features in the
stacked spectra that can give further insight into the gas covering
fraction and the distribution of metals and dust between the neutral
and ionized ISM.

\subsection{Ly$\alpha$ Escape Fractions}
\label{sec:lyaescape}

Given the correspondence between Ly$\alpha$ strength and
reddening/covering fraction noted in previous studies (e.g.,
\citealt{shapley03, kornei10, wofford13, jones13, rivera15,
  trainor15}), we calculated the absolute Ly$\alpha$ escape fraction,
\begin{eqnarray}
f({\rm Ly}\alpha)_{\rm abs} & = & \frac{L({\rm Ly}\alpha)_{\rm obs}}{L({\rm Ly}\alpha)_{\rm int}},
\end{eqnarray}
where $L({\rm Ly}\alpha)_{\rm obs}$ and $L({\rm Ly}\alpha)_{\rm int}$
are the average observed and intrinsic Ly$\alpha$ line luminosities,
respectively, for the galaxies contributing to each composite shown in
Figure~\ref{fig:confint}.  The observed line luminosity for each
spectral stack was calculated as follows.  We first redshifted the
composite to the mean redshift of objects contributing to that
composite.  The spectrum was then converted to $f_{\lambda}$ units.
The observed Ly$\alpha$ line luminosity was calculated by directly
integrating the flux over just the Ly$\alpha$ emission profile (i.e.,
from the points where the emission line intersects with the underlying
Ly$\alpha$ absorption profile) using the IRAF task {\em splot}.
Naturally, this calculation will not account for the potentially large
fraction of Ly$\alpha$ (up to $80\%$; \citealt{steidel11}) that may be
resonantly scattered into a diffuse halo (see discussion below).

The intrinsic Ly$\alpha$ line luminosity was estimated from the
star-formation rate (SFR).  The SFR was calculated using the UV
luminosity density at $1500$\,\AA, corrected for dust based on the
average $\ebmv$ of objects contributing to each composite and assuming
the \citet{reddy15} attenuation curve,\footnote{These are similar to
  the dust correction factors obtained with the \citet{calzetti00}
  attenuation curve.} and converted to SFR assuming the
\citet{kennicutt98} relation for a \citet{salpeter55} IMF.  The
intrinsic Ly$\alpha$ luminosity was then estimated from the SFR using
a modified form of the relation given in \citet{kennicutt98}, where we
assumed that the intrinsic ratio of the Ly$\alpha$-to-H$\alpha$ line
intensities is $f({\rm Ly}\alpha)/f({\rm H}\alpha) \simeq 8.7$ for
Case B recombination and $T_{\rm e} = 10^4$\,K.  This yields the following
relationship between SFR and Ly$\alpha$ line luminosity, again assuming
a \citet{salpeter55} IMF:
\begin{eqnarray}
L({\rm Ly}\alpha)_{\rm int}\,[{\rm ergs}\,{\rm s}^{-1}] = \frac{{\rm SFR}\,[M_\odot\,{\rm yr}^{-1}]}{9.1\times 10^{-43}}.
\end{eqnarray}

The measurement error in the mean escape fraction was estimated by
perturbing the composite many times according to its error spectrum,
and calculating the observed and intrinsic Ly$\alpha$ luminosities for
each of these realizations.  Note that the observed Ly$\alpha$ line
luminosity and SFR are calculated using the spectrum itself; we made
no corrections for slit loss, and thus the escape fractions derived
here assume that the Ly$\alpha$ and UV continuum fluxes arise from
spatially coincident regions.  The resulting $f({\rm Ly}\alpha)_{\rm
  abs}$ are indicated for each composite in Figure~\ref{fig:confint}.

\begin{figure}
\epsscale{0.90}
\plotone{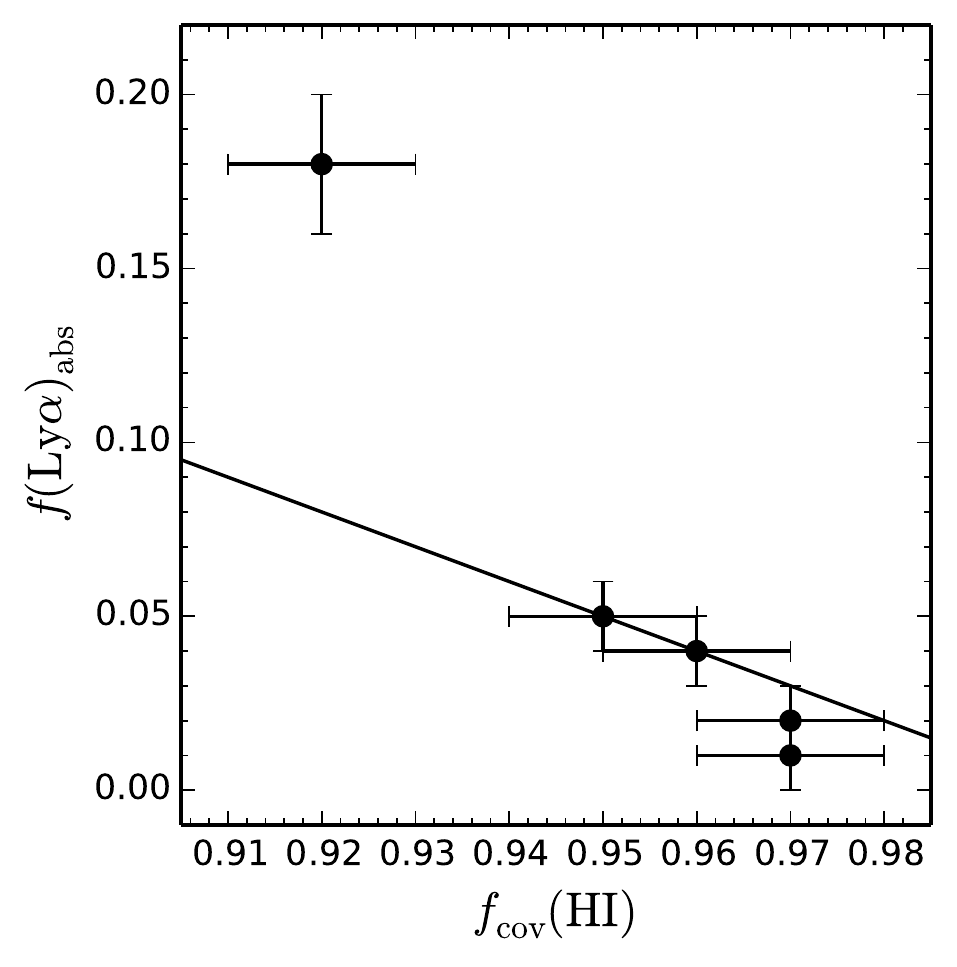}
\caption{Variation of the absolute escape fraction of Ly$\alpha$
  photons ($f({\rm Ly}\alpha)_{\rm abs}$) with neutral gas covering
  fraction, $f_{\rm cov}(\hi)$.  Errors were determined by perturbing
  the composite spectra many times according to the error spectra and
  recalculating $f({\rm Ly}\alpha)_{\rm abs}$ and $f_{\rm cov}(\hi)$.
  The solid line shows the relation $f({\rm Ly}\alpha)_{\rm abs} =
  1-f_{\rm cov}(\hi)$, assuming that the Ly$\alpha$ photons escape
  through clear sightlines in the ISM.  Note that the Ly$\alpha$
  escape fractions shown here do not include the potentially large
  fraction of Ly$\alpha$ photons that may be resonantly scattered out
  of the slit aperture (see text).}
\label{fig:lyafrac}
\end{figure}

Not surprisingly, $f({\rm Ly}\alpha)_{\rm abs}$ correlates with
continuum reddening (and hence with $f_{\rm cov}(\hi)$;
Figure~\ref{fig:lyafrac}; see also \citealt{steidel10}) with a Pearson
correlation coefficient of $\rho = -0.97$ and a probability of a null
correlation of $p=0.005$.  For four of the five $\ebmv$ bins, the
absolute escape fraction of Ly$\alpha$ photons agrees well with our
expectation if such photons only escaped through clear sightlines in
the ISM; namely, $f({\rm Ly}\alpha)_{\rm abs} \approx 1 - f_{\rm
  cov}(\hi)$ (Figure~\ref{fig:lyafrac}).  The bluest bin of $\ebmv$
exhibits a larger $f({\rm Ly}\alpha)_{\rm abs}$ than this expectation.
This may suggest that a significant fraction of the Ly$\alpha$ flux
may be resonantly scattered into the spectroscopic slit aperture
and/or may arise from Ly$\alpha$ photons scattering out of resonance,
though we note that the kinematic profile of the Ly$\alpha$ line for
the galaxies in the bluest $\ebmv$ bin is not substantially different
than that of the other bins.  Moreover, it is unclear why the fraction
of Ly$\alpha$ photons scattered into the spectroscopic slit would be
much larger for galaxies with bluer $\ebmv$.  A more detailed
assessment of {\em how} Ly$\alpha$ photons escape these galaxies (as
judged through the kinematic profile of the line) must await more
precise systemic redshifts derived from nebular emission lines.  

In the context of Figure~\ref{fig:cartoon}, one might expect the
fraction of Ly$\alpha$ emission exiting through low column density
sightlines to perhaps correlate more directly with $f_{\rm cov}(\hi)$
than the diffuse Ly$\alpha$ component; the latter is not accounted for
in our calculation.  However, based on the results of the previous
section, both $\nhi$ and $\ebmv_{\rm los}$ increase in tandem with
$f_{\rm cov}(\hi)$.  Hence, Ly$\alpha$ photons that diffuse far from
the sites of star formation face an increasing probability of
absorption by dust with increasing $f_{\rm cov}(\hi)$, implying that
the fraction of Ly$\alpha$ scattered into a diffuse halo and
ultimately exiting the ISM/CGM of the galaxy is also likely to
correlate with $f_{\rm cov}(\hi)$.  This inference could have also
been deduced from the results of \citet{steidel11}, who find
that---when considering resonantly scattered Ly$\alpha$---the
attenuation of Ly$\alpha$ photons is consistently larger than that of
the UV continuum, and the fraction of the total Ly$\alpha$ produced
from star formation that diffuses into a halo correlates inversely
with the spectroscopic Ly$\alpha$ flux.  At any rate, the computed
$f({\rm Ly}\alpha)_{\rm abs}$ imply that the selection of galaxies
based on strong Ly$\alpha$ emission will also tend to isolate those
with lower gas covering---and higher LyC escape---fractions.  In
accord with expectations, the conditions that give rise to strong
Ly$\alpha$ emission also facilitate the escape of LyC radiation.

\subsection{Interstellar Absorption Lines}
\label{sec:isabslines}

\begin{figure*}[!t]
\epsscale{1.00}
\plotone{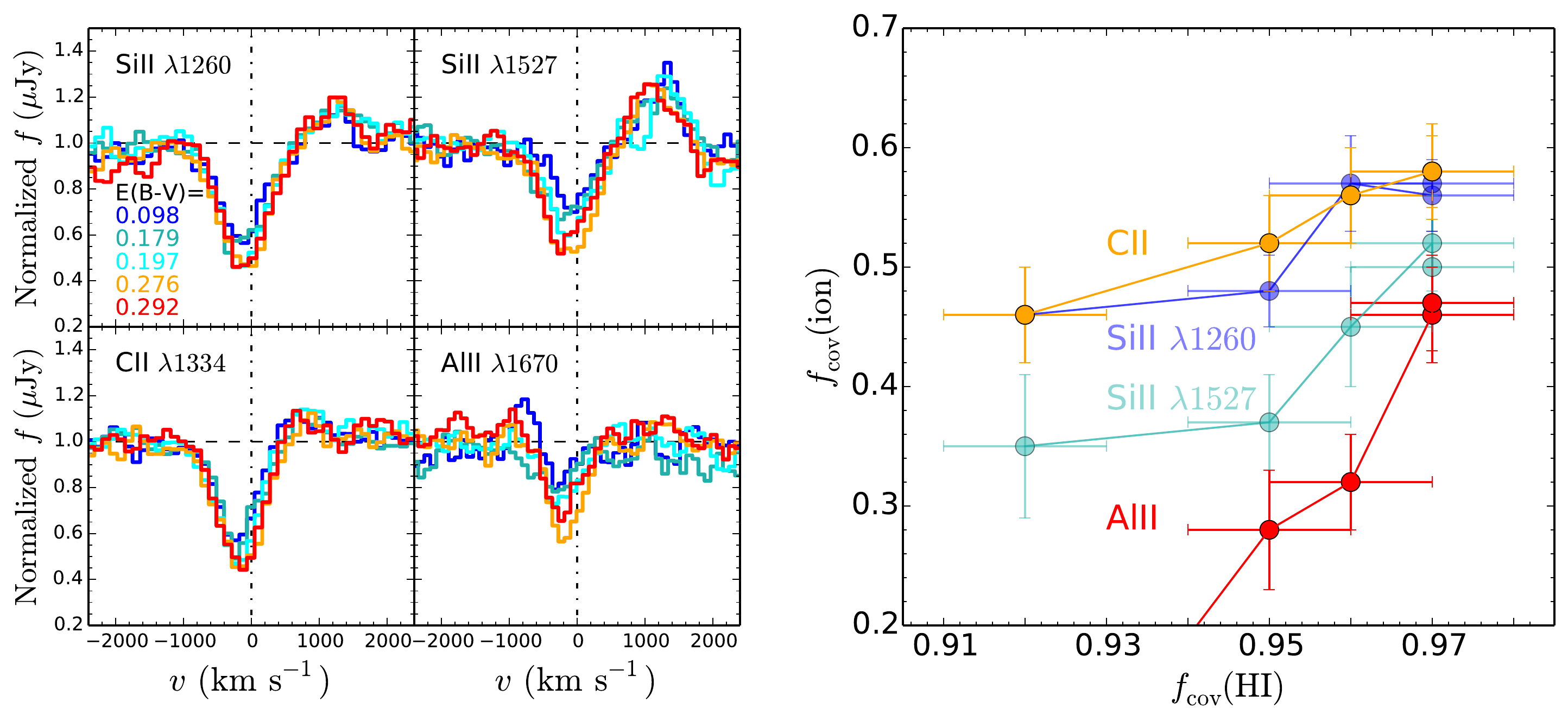}
\caption{{\em Left:} Normalized composite spectra in different bins of
  $\ebmv$ for several strong low-ionization interstellar absorption
  lines, including \ion{Si}{2}\,$\lambda 1260$, \ion{Si}{2}\,$\lambda
  1527$, \ion{C}{2}\,$\lambda 1334$, and \ion{Al}{2}\,$\lambda 1670$.
  The normalized composites were constructed by dividing the spectra
  shown in Figure~\ref{fig:confint} by their best-fitting models.  In
  all cases, the bulk of the absorbing gas is blueshifted by $\approx
  300$\,km\,s$^{-1}$, similar to that observed for the $\hi$ lines.
  Additionally, the two \ion{Si}{2} transitions at $1260$ and
  $1527$\,\AA\, are saturated and suggest a covering fraction of
  \ion{Si}{2}-enriched material of between $35\%$ to $60\%$, and
  increasing with continuum reddening (see also \citealt{shapley03}).
  The other low-ionization lines indicate similar covering fractions,
  assuming that these transitions are also optically thick.  {\em
    Right:} The derived covering fractions for \ion{Si}{2}\,$\lambda
  1260$ ({\em blue points}), \ion{Si}{2}\,$\lambda 1527$ ({\em cyan
    points}), \ion{C}{2}\,$\lambda 1334$ ({\em orange points}), and
  \ion{Al}{2}\,$\lambda 1670$ ({\em red points}), versus $f_{\rm
    cov}(\hi)$.  The errors in the covering fraction were deduced by
  perturbing the normalized composite spectra by the normalized error
  spectra many times and, each time, locating the point of maximum
  absorption (minimum flux).  The dispersion in the minimum fluxes
  then determined the errors in the covering fractions for each of the
  composites and ionic species.}
\label{fig:lowions}
\end{figure*}

Given the prevalent use of low-ionization interstellar absorption
lines as proxies for the neutral gas covering fraction in
high-redshift galaxies (e.g., \citealt{shapley03, jones13, trainor15,
  alexandroff15}), it is instructive to directly compare the covering
fractions deduced from the low-ionization lines with those computed
from the depth of the $\hi$ lines.  To accomplish this, the five
composites shown in Figure~\ref{fig:confint} were divided by their
respective best-fitting models to produce normalized composites.  This
method of normalization has the advantage of removing any stellar
absorption component (e.g., \ion{C}{2}\,$\lambda 1334$) that may
affect equivalent width measurements of the interstellar absorption
lines.  The normalized composites were transformed into velocity space
around the \ion{Si}{2}\,$\lambda 1260$, \ion{Si}{2}\,$\lambda 1527$,
\ion{C}{2}\,$\lambda 1334$, and \ion{Al}{2}\,$\lambda 1670$ IS lines.
These include the strongest and most commonly-used IS lines observed
in the rest-frame UV spectra of high-redshift galaxies.

These features are shown in Figure~\ref{fig:lowions}, with the
different colors indicating the normalized spectrum in each bin of
$\ebmv$.  If the two \ion{Si}{2} transitions at $1260$ and
$1527$\,\AA\, are optically thin, then the ratio of their
equivalent widths will be $W_{\rm 1260}/W_{\rm 1527} \ga 6$.  The
observed ratio is $W_{\rm 1260}/W_{\rm 1527} \simeq 1$, implying that
the lines are saturated and hence their depths are sensitive to the
covering fraction of \ion{Si}{2}-enriched material.  
Other strong low-ionization IS absorption features are shown in
Figure~\ref{fig:lowions}, including \ion{C}{2}\,$\lambda 1334$, and
\ion{Al}{2}\,$\lambda 1670$ lines.  The similarity in the velocity
widths and depths of these additional lines to those of the saturated
\ion{Si}{2} transitions discussed above suggests that the former may
be also arise from the same gas as the latter.  If this is the case,
then collectively the depths of the low-ionization lines indicate a
covering fraction that rises from $f_{\rm cov}(\text{ion})\approx
0.30$ to $0.65$ proceeding from the bluest to reddest $\ebmv$ bin (see
also \citealt{shapley03}).  The errors in $f_{\rm cov}(\text{ion})$
were calculated by perturbing many times the normalized composite
spectra by the normalized error spectra, and measuring the depth of
the absorption lines for each of these realizations.  The dispersion
in the absorption line depths yielded the measurement uncertainty in
$f_{\rm cov}(\text{ion})$, but of course does not include systematic
errors associated with the specific model for the continuum level.

The covering fractions of the low-ionization species are correlated
with, but are systematically lower than, those derived from the $\hi$
lines (lower right panel of Figure~\ref{fig:lowions}).  There are
several possible explanations for why the low-ionization lines suggest
lower covering fractions.  First, if most of the metals inhabit narrow
and unresolved absorption systems, then the depth of the metal lines
will underpredict the covering fraction.  Likewise, if the
metal-enriched gas has a substantial velocity gradient across the face
of the galaxy, then the derived covering fractions may be lower limits
(e.g., see discussion in Section~\ref{sec:evidencefornonunity}).
While we cannot definitively rule out either of these possibilities,
the similarity in the blueshift ($\approx 200$\,km\,s$^{-1}$) and
velocity spread (FWHM$\approx 800$\,km\,s$^{-1}$; e.g., compare
Figure~\ref{fig:lowions} to the right panel of
Figure~\ref{fig:subtract}) of the low-ionization and $\hi$ lines
suggests that the two components are kinematically coupled, where the
former covers a smaller portion of the galaxy's continuum.  Such a
configuration may arise if dense and metal-enriched pockets of gas are
embedded within the outflowing $\hi$.  Consequently, some substantial
fraction of the outflowing gas may be metal-poor.


\begin{figure*}[!t]
\epsscale{1.00}
\plotone{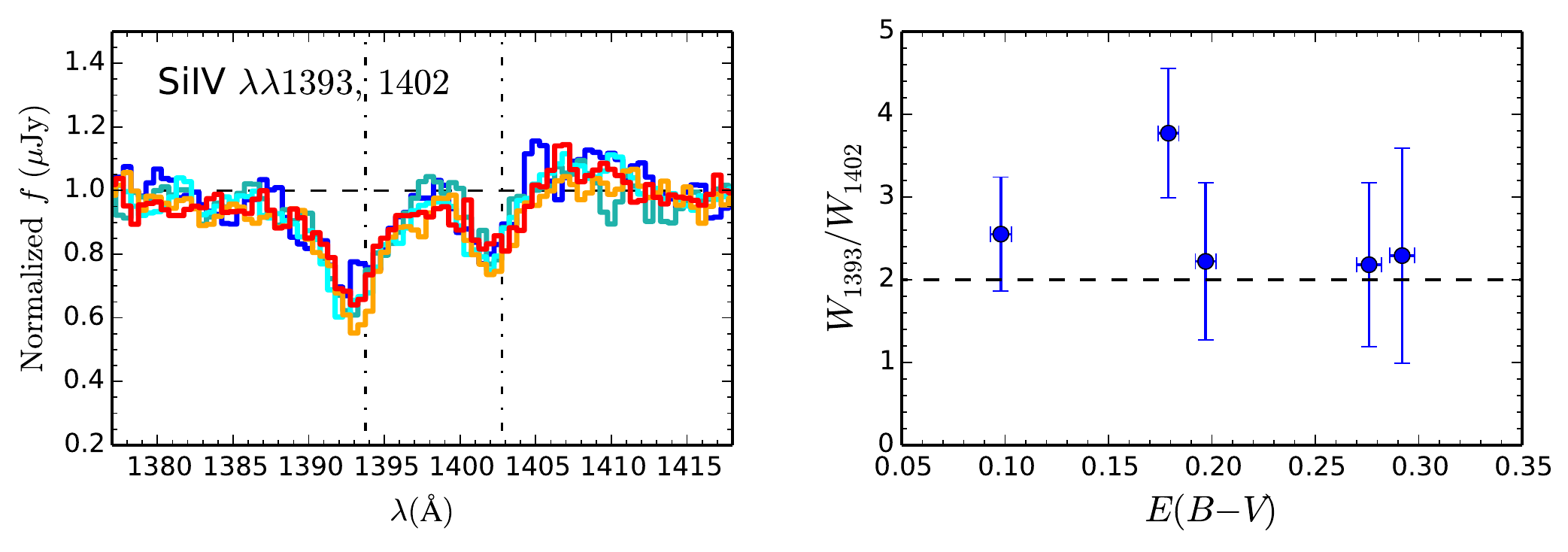}
\caption{{\em Left:} Normalized composite spectra in different bins of
  $\ebmv$ (same color coding as in Figure~\ref{fig:lowions}) for the
  high-ionization \ion{Si}{4}\,$\lambda\lambda 1393, 1402$ lines. {\em
    Right:} The ratios of the equivalent widths of the two lines as a
  function of $\ebmv$ for the five composites suggest an optical depth
  $\tau(\text{\ion{Si}{4}}) \simeq 1$.}
\label{fig:highions}
\end{figure*}

Additionally, we note that $f_{\rm cov}({\rm ion})$ increases by a
factor of $\approx 1.5$ in response to a small ($\approx 5\%$) increase in
$f_{\rm cov}(\hi)$.  This behavior suggests that as $f_{\rm cov}(\hi)$ increases, 
a larger fraction of the neutral gas is enriched with \ion{Si}{2}, implying
a corresponding increase in the average dust-to-gas ratio of the outflowing gas.
%
Notwithstanding the fact that the line-of-sight reddening will be
sensitive to the total column of dust (including that of the ionized
gas), an increase in the dust-to-gas ratio of the outflowing gas for
objects with redder continuum would result in a strong correlation
between $\ebmv_{\rm los}$ and $\ebmv$, consistent with what we
actually observe (left panel of Figure~\ref{fig:covfrac}).

Figure~\ref{fig:highions} shows the profiles of the two
\ion{Si}{4}\,$\lambda\lambda 1393, 1402$ transitions for each bin of
reddening.  As the ionization potential for \ion{Si}{4} ($q=33.5$\,eV)
is much higher than that required for H-ionization, the associated
transitions arise from H-ionized gas.  The lines are blueshifted by
$\approx 160$\,km\,s$^{-1}$, slightly smaller than the blueshifts
derived from the low-ionization and $\hi$ lines, though a more
detailed comparison of the kinematics of the neutral, and low- and
high-ionization gas will rely on future precise measurements of the
systemic redshifts of galaxies in our sample.  The ratio of the
equivalent widths of the \ion{Si}{4} lines is $W_{\rm 1393}/W_{\rm
  1402} \approx 2$, indicating that the absorption is consistent
within the errors of being on the linear part of the curve of growth.
As the depths of the lines do not correlate obviously with the
reddening of the continuum,
we surmise that the outflowing ionized gas has an optical depth $\tau
\simeq 1$.  Note that the errors in $W_{\rm 1393}/W_{\rm 1402}$ do not
rule the possibility of the lines being saturated, in which case the
covering fraction of the ionized material is $f_{\rm
  cov}(\text{\ion{Si}{4}}) \ga 30\%$.  More generally, the presence of
these high-ionization IS absorption lines suggests an ionized medium
mixed with metals and dust.

\subsection{Summary of Results from Ly$\alpha$ and Interstellar Absorption Lines}

To summarize, we find that the absolute escape fraction of Ly$\alpha$
photons, $f({\rm Ly}\alpha)_{\rm abs}$, inversely correlates with
$f_{\rm cov}(\hi)$, suggesting that the neutral gas covering fraction
plays a significant role in modulating the escape of Ly$\alpha$
photons.  The low-ionization interstellar doublet
\ion{Si}{2}\,$\lambda\lambda 1260, 1527$ is saturated, exhibits
similar absorption depths and velocity profiles to those of other
low-ionization IS absorption lines, and suggests a covering fraction
of low-ionization material, $f_{\rm cov}(\text{ion})$, that increases
with continuum reddening (and $f_{\rm cov}(\hi)$) from $\approx 0.30$
for the bluest galaxies in our sample to $\approx 0.65$ for the
reddest ones.  However, $f_{\rm cov}(\text{ion})$ is systematically
lower than $f_{\rm cov}(\hi)$, while the low-ionization lines have
similar blueshifts and velocity profiles as the $\hi$ lines,
suggesting that galaxy outflows consist in part of a neutral medium
embedded with clumps of metal-enriched gas.  Moreover, the rate of
increase of $f_{\rm cov}(\text{ion})$ relative to $f_{\rm cov}(\hi)$
implies that the average column of dust increases with covering
fraction, consistent with our observation that the line-of-sight
reddening correlates with continuum reddening.  These results imply
that a significant fraction of the neutral medium may be chemically
pristine, and that low-ionization interstellar absorption lines may
severely underpredict the neutral gas covering fraction.  A similar
analysis for the high-ionization absorption lines, namely
\ion{Si}{4}\,$\lambda\lambda 1393, 1402$, suggests that the
high-ionization gas is not optically thick and has $f_{\rm
  cov}(\text{\ion{Si}{4}}) \ga 30\%$.

\section{CONNECTION BETWEEN CONTINUUM REDDENING AND ESCAPE OF IONIZING PHOTONS}
\label{sec:escape}

\subsection{Gas Covering and Lyman Continuum Escape Fractions}

Lyman continuum (LyC) photons may be extinguished by dust absorption
or through hydrogen photoelectric absorption.  In the case of dust
absorption, our new measurement of the far-UV shape of the dust curve
at high redshift, presented in Paper~I, implies a lower attenuation of
LyC photons than those inferred from simple extrapolations of the
\citet{calzetti00} and \citet{reddy15} attenuation curves.
Alternatively, the high column densities ($\lognhi \ga 20.3$) inferred
from the fitting presented in Section~\ref{sec:specfitting} suggest
that the covered portions of a galaxy are optically thick to LyC
radiation, with implied optical depths at $900$\,\AA\, of
$\tau_{900}\simeq 1150$.  Thus, for the average galaxy, photoelectric
absorption, rather than dust, dominates the depletion of LyC photons.
Similar results have been suggested by simulations of escaping
ionizing radiation from high-redshift galaxies \citep{gnedin08,
  razoumov10, ma16}.  Accordingly, in the geometry illustrated in
Figure~\ref{fig:cartoon}, LyC photons will only escape from dust-free
holes in the ISM.  The very high optical depth of the covering neutral
gas suggests that Ly$\alpha$ photons will also predominantly escape
from regions cleared of dust and gas, an expectation that is verified
by the strong correlation between $f({\rm Ly}\alpha)_{\rm abs}$ and
reddening/covering fraction (Section~\ref{sec:lyaescape}).

LyC extinction may be a more significant effect in galaxies where the
radiation field is sufficient to completely ionize the hydrogen (i.e.,
corresponding to a ``density-bounded'' case; see
\citealt{zackrisson13}), but such galaxies must be relatively rare in
our sample given the apparent frequency of high column density neutral
gas as reflected in the composite spectra.  We can connect the gas
covering fraction to the absolute escape fraction of LyC photons,
where the latter is commonly written as:
\begin{eqnarray}
f({\rm LyC})_{\rm abs} & = & \exp[-\tau_{\rm ISM}({\rm LyC})] \times \nonumber \\
& & 10^{-0.4k({\rm LyC})\ebmv}.
\end{eqnarray}
The first term in this expression accounts for the opacity of the ISM,
where $\tau_{\rm ISM}({\rm LyC})$ is the optical depth at LyC
wavelengths.  The second term accounts for the obscuration of LyC
photons by dust, where $\ebmv$ is the reddening and $k({\rm LyC})$ is
the dust attenuation curve evaluated at LyC wavelengths.  In the case
where the dust and neutral gas completely cover the continuum, the
opacity of the ISM can be combined with the dust attenuation curves
calculated in Paper~I to compute $f({\rm LyC})_{\rm abs}$.

In the case of incomplete covering, the absolute escape fraction of LyC
photons can be written as:
\begin{eqnarray}
f({\rm LyC})_{\rm abs} & = & f_{\rm cov}(\hi)\exp[-\tau_{\rm ISM}^{\rm cov}({\rm LyC})] \times \nonumber \\
& & 10^{-0.4k({\rm LyC})\ebmv^{\rm cov}_{\rm los}} + \nonumber \\
& & (1-f_{\rm cov}(\hi))\exp[-\tau_{\rm ISM}^{\rm uncov}({\rm LyC})] \times \nonumber \\ 
\nonumber \\
& & 10^{-0.4k({\rm LyC})\ebmv^{\rm uncov}_{\rm los}}, 
\label{eq:twoterms}
\end{eqnarray}
where the first and second terms correspond to the covered and
uncovered portions contributing to the galaxy spectrum, respectively.
In this expression, $k({\rm LyC})$ is a line-of-sight extinction curve
evaluated at LyC wavelengths, and $\ebmv^{\rm cov}_{\rm los}$ and $\ebmv^{\rm
  uncov}_{\rm los}$ indicate the line-of-sight reddening appropriate to the
covered and uncovered continuum, respectively.

Similarly, the absolute escape fraction of UV continuum photons
(e.g., typically measured between $1500$ and $1700$\,\AA) is
\begin{eqnarray}
f({\rm UV})_{\rm abs} & = & 10^{-0.4k({\rm UV})\ebmv},
\label{eq:fuv1}
\end{eqnarray}
where $k({\rm UV})$ refers to the attenuation curve relevant for the
stellar continuum, assuming a unity covering fraction of
dust.\footnote{Note that we can also write $f({\rm UV})_{\rm abs}$ as
  two terms that refer to the covered and uncovered portions of the
  continuum, analogous to Equation~\ref{eq:twoterms}:
\begin{eqnarray}
f({\rm UV})_{\rm abs} & = & f_{\rm cov}(\hi)\times 10^{-0.4k({\rm UV})\ebmv^{\rm cov}_{\rm los}} + \nonumber \\
& & (1-f_{\rm cov}(\hi))\times 10^{-0.4k({\rm UV})\ebmv^{\rm uncov}_{\rm los}},
\label{eq:fuv2a}
\end{eqnarray}
which, in the case of low dust column density for the uncovered portion,
can be written as
\begin{eqnarray}
f({\rm UV})_{\rm abs} & \approx & f_{\rm cov}(\hi)\times 10^{-0.4k({\rm UV})\ebmv^{\rm cov}} + \nonumber \\
& & (1-f_{\rm cov}(\hi)),
\label{eq:fuv2b}
\end{eqnarray}
where $k({\rm UV})$ now refers to a line-of-sight extinction curve.
As the color excess is measured most conveniently assuming a unity
covering fraction of dust with a stellar attenuation curve, these
equations are of less practical use than Equation~\ref{eq:fuv1}.}

These equations are cast most conveniently in terms of the ``relative''
escape fraction defined as
\begin{eqnarray}
f_{\rm rel} & = & \frac{f({\rm LyC})_{\rm abs}}{f({\rm UV})_{\rm abs}}
\label{eq:frel1}
\end{eqnarray}
\citep{steidel01}.  The relationship between the relative escape
fraction and the {\em observed} ratio of LyC-to-UV flux density is
\begin{eqnarray}
f_{\rm rel} & = & \frac{S({\rm LyC})_{\rm obs}/S({\rm LyC})_{\rm int}}{S({\rm UV})_{\rm obs}/S({\rm UV})_{\rm int}}\times \nonumber \\
& & \exp[\tau_{\rm IGM}({\rm LyC})]\times \exp[\tau_{\rm CGM}({\rm CGM})],
\label{eq:frel2}
\end{eqnarray}
where $S({\rm LyC})_{\rm obs}$ and $S({\rm UV})_{\rm obs}$ are the
observed flux densities of LyC and UV continuum photons, respectively;
$S({\rm LyC})_{\rm int}$ and $S({\rm UV})_{\rm int}$ are the intrinsic
flux densities of LyC and UV continuum photons, respectively, as
typically estimated from stellar population synthesis models; and the
last two terms indicate the further depletion of LyC photons due to
the opacity of the IGM and CGM.

Combining Equations~\ref{eq:fuv1}, \ref{eq:frel1}, and \ref{eq:frel2}
yields the following expression for the observed LyC-to-UV flux
density ratio:
\begin{eqnarray}
\left[ \frac{S({\rm LyC})}{S({\rm UV})}\right]_{\rm obs} & = & \left[ \frac{S({\rm LyC})}{S({\rm UV})}\right]_{\rm int} \times \nonumber \\
& & \exp[-\tau_{\rm IGM}({\rm LyC})] \times \exp[-\tau_{\rm CGM}({\rm LyC})] \times \nonumber \\
& & \left[\frac{f({\rm LyC})_{\rm abs}}{10^{-0.4k({\rm UV})\ebmv}}\right],
\label{eq:final1}
\end{eqnarray}
where $f({\rm LyC})_{\rm abs}$ is given by Equation~\ref{eq:twoterms}.

For the average $z\sim 3$ galaxy in our sample, the first term of
Equation~\ref{eq:twoterms} is negligible owing to the large neutral
gas column density and hence large value of $\tau_{\rm ISM}^{\rm
  cov}$, as noted above.  Hence, if the uncovered portions of the
galaxy continuum are characterized by very low gas and dust column
densities, as would be expected in the simple geometry of
Figure~\ref{fig:cartoon}, then the absolute escape fraction of LyC
photons is simply
\begin{eqnarray}
f({\rm LyC})_{\rm abs} & \approx & 1-f_{\rm cov}(\hi).
\label{eq:lycabs}
\end{eqnarray}
We noted in Section~\ref{sec:evidencefornonunity} that $f({\rm
  LyC})_{\rm abs}$ may be lower than that indicated by
Equation~\ref{eq:lycabs}.  We discuss this further in
Section~\ref{sec:applications}.  In the present context,
Equation~\ref{eq:final1} becomes
\begin{eqnarray}
\left[\frac{S({\rm LyC})}{S({\rm UV})}\right]_{\rm obs} & = & \left[\frac{S({\rm LyC})}{S({\rm UV})}\right]_{\rm int} \times \nonumber \\
& & \exp[-\tau_{\rm IGM}({\rm LyC})] \times \exp[-\tau_{\rm CGM}({\rm LyC})] \times \nonumber \\
& & \left[\frac{1-f_{\rm cov}(\hi)}{10^{-0.4k({\rm UV})\ebmv}}\right].
\label{eq:final1b}
\end{eqnarray}
In practice, $f_{\rm cov}(\hi)$ is difficult to constrain in the
absence of direct spectroscopy of the $\hi$ absorption lines---e.g.,
as noted in Section~\ref{sec:isabslines}, neutral gas covering
fractions deduced from low-ionization IS absorption lines may severely
underpredict those derived from the $\hi$ lines.  Thus, the utility of
an empirical relation between $f_{\rm cov}(\hi)$ and $\ebmv$ becomes
obvious: the continuum reddening is relatively easy to measure, even
with just a few broadband photometric filters covering the rest-frame
UV.  As discussed in Section~\ref{sec:correlations} and shown in
Figure~\ref{fig:covfrac} and Table~\ref{tab:covfracfits}, we find a
significant correlation between $f_{\rm cov}(\hi)$ and $\ebmv$.  Thus,
a suitable functional relationship between these two variables can be
used to cast Equations~\ref{eq:twoterms} and \ref{eq:final1b} in terms
of $\ebmv$.  The functional form relating $f_{\rm cov}(\hi)$ and
$\ebmv$ is the subject of the next subsection.

\subsection{Functional Form for the Gas Covering Fraction in Terms of Continuum Reddening}

The middle and right panels of Figure~\ref{fig:covfrac} show the
relationship between $f_{\rm cov}(\hi)$ and $\ebmv$.  The Pearson
correlation coefficients when we assume the SMC and Milky Way
extinction curves in computing $f_{\rm cov}(\hi)$ are $\rho=0.90$ and
$0.87$, respectively, with probabilities for no correlation between
$f_{\rm cov}(\hi)$ and $\ebmv$ of $p=0.04$ and $0.05$, respectively.
To uncover the functional form of $f_{\rm cov}(\hi)$ vs. $\ebmv$, we
reddened the intrinsic template assuming different values of $\ebmv$
and the updated and original forms of the \citet{reddy15} and
\citet{calzetti00} attenuation curves.  For each of these reddened
templates, we determined which combination of the covering fraction
and line-of-sight reddening---assuming either the SMC or Milky Way
extinction curves---provided the best match in the ``Composite
$\ebmv$'' wavelength windows listed in Table~\ref{tab:waves}, and one
additional window covering the LyC region: $\lambda=800-900$\,\AA.
The model expectations for various assumptions of the reddening and
line-of-sight dust curves are shown in the middle and right panels of
Figure~\ref{fig:covfrac}.  For clarity, not all of the combinations of
attenuation/extinction curves are shown, as they all result in a
similar functional behavior between $f_{\rm cov}(\hi)$ and $\ebmv$.

As expected, the covering fraction asymptotes to zero and
unity for small and large values of the continuum reddening,
respectively.  Motivated by this behavior, we adopted the following
functional form for $f_{\rm cov}(\hi)$ vs. $\ebmv$:
\begin{eqnarray}
f_{\rm cov}(\hi) = 1 - \exp[a\times\ebmv^b],
\label{eq:functional}
\end{eqnarray}
where $a$ and $b$ are free parameters.  Table~\ref{tab:covfracfits}
summarizes the free parameters obtained by fitting the {\em measured}
values of $f_{\rm cov}$ and $\ebmv$ for the updated \citet{reddy15}
attenuation curve and the SMC extinction curve applicable to the
continuum and line-of-sight reddening, respectively.

To compute these fits, we randomly perturbed the color excess and
covering fraction measurements according to their respective errors
and fit each of these realizations with Equation~\ref{eq:functional}.
The average and standard deviation in the best-fit parameters among
these realizations are the values reported in
Table~\ref{tab:covfracfits}.  The $68\%$ confidence intervals in the
fits were computed as the $1$\,$\sigma$ dispersion in the fits to the
realizations at each value of $\ebmv$.  These confidence intervals do
not vary significantly over the range of $\ebmv$ probed in our study
for different assumptions of the attenuation curve, as the fits are
forced to go through the measured data.

Finally, we can combine Equations~\ref{eq:final1b} and
\ref{eq:functional} to obtain an expression for the observed LyC-to-UV
flux density ratio in terms of $\ebmv$ for the geometry illustrated in
Figure~\ref{fig:cartoon}:
\begin{eqnarray}
\Biggl\langle \frac{S({\rm LyC})}{S({\rm UV})}\Biggl\rangle_{\rm obs} & = & \Biggl\langle \frac{S({\rm LyC})}{S({\rm UV})}\Biggl\rangle_{\rm int} \times \nonumber \\
& & \langle \exp[-\tau_{\rm IGM}({\rm LyC})] \times \exp[-\tau_{\rm CGM}({\rm LyC})] \rangle \times \nonumber \\
& & \left[\frac{\exp[a\times\langle\ebmv\rangle^b]}{10^{-0.4k({\rm UV})\langle\ebmv\rangle}}\right].
\label{eq:final2}
\end{eqnarray}
We have written this equation in terms of averages to emphasize that
the relation applies for an ensemble of galaxies.  Adopting values for
the IGM and CGM opacities appropriate for the redshifts of interest
allows us to use Equations~\ref{eq:twoterms}, \ref{eq:final1}, and
\ref{eq:final2} in several useful ways, as we discuss in
Section~\ref{sec:applications}.

\subsection{Extending the Model of LyC Escape to Bluer Galaxies}
\label{sec:extendmodels}


To uncover the extent to which the relationship between $[S({\rm
    LyC})/S({\rm UV})]_{\rm obs}$ and $\ebmv$ may change over a larger
range of $\ebmv$ than accessible in the current sample, we examined
how several of the other factors relevant in the escape of LyC
photons, namely $\nhi$ and $\ebmv_{\rm los}$, vary with $\ebmv$.  As
noted in Section~\ref{sec:correlations} and shown in
Figure~\ref{fig:covfrac} and Table~\ref{tab:covfracfits}, several
significant correlations are found between these factors and $\ebmv$.
We fit $\log\nhi$ vs. $\log \ebmv$ and $\log\ebmv_{\rm los}$ vs. $\log
\ebmv$ with linear functions.  The best-fitting intercepts and slopes,
and their uncertainties, are summarized in
Table~\ref{tab:covfracfits}.  
We used these linear fits to estimate $\nhi$ and $\ebmv_{\rm los}$ for
galaxies with $\ebmv \la 0.09$, and used Equation~\ref{eq:twoterms} to
calculate $f({\rm LyC})_{\rm abs}$, and Equation~\ref{eq:final1} to
calculate $[S({\rm LyC})/S({\rm UV})]_{\rm obs}$.  In the next
section, we examine how $[S({\rm LyC})/S({\rm UV})]_{\rm obs}$ varies
with continuum reddening.

\subsection{Practical Applications of the Model of LyC Escape}

\subsubsection{Calculating the Models}

To calculate the predicted $\langle S({\rm LyC})/S({\rm
  UV})\rangle_{\rm obs}$ from our model, we first calculated the
expected opacities of the IGM and CGM by considering the neutral gas
column density distribution in the Ly$\alpha$ forest at $z\sim
2.0-2.8$, as well as the excess $\hi$ absorption around $\sim
L^{\ast}$ galaxies at the same redshifts, as quantified in
\citet{rudie13}.  Using 1000 random realizations of these column
density distributions, we calculate the attenuation of $900$\,\AA\,
photons to be $\langle \exp[-\tau_{\rm IGM}(900)] \times
\exp[-\tau_{\rm CGM}(900)]\rangle = 0.37$ for galaxies at $z=3.05$,
corresponding to the mean redshift of the 121 galaxies that contribute
to the composite at $\lambda < 1150$\,\AA.  The dispersion in the
transmissivity among the 1000 realizations at this wavelength is
$\approx 0.33$, implying that the error in the mean transmissivity for
the 121 galaxies is $\approx 8.1\%$.

\begin{figure}
\epsscale{1.15}
\plotone{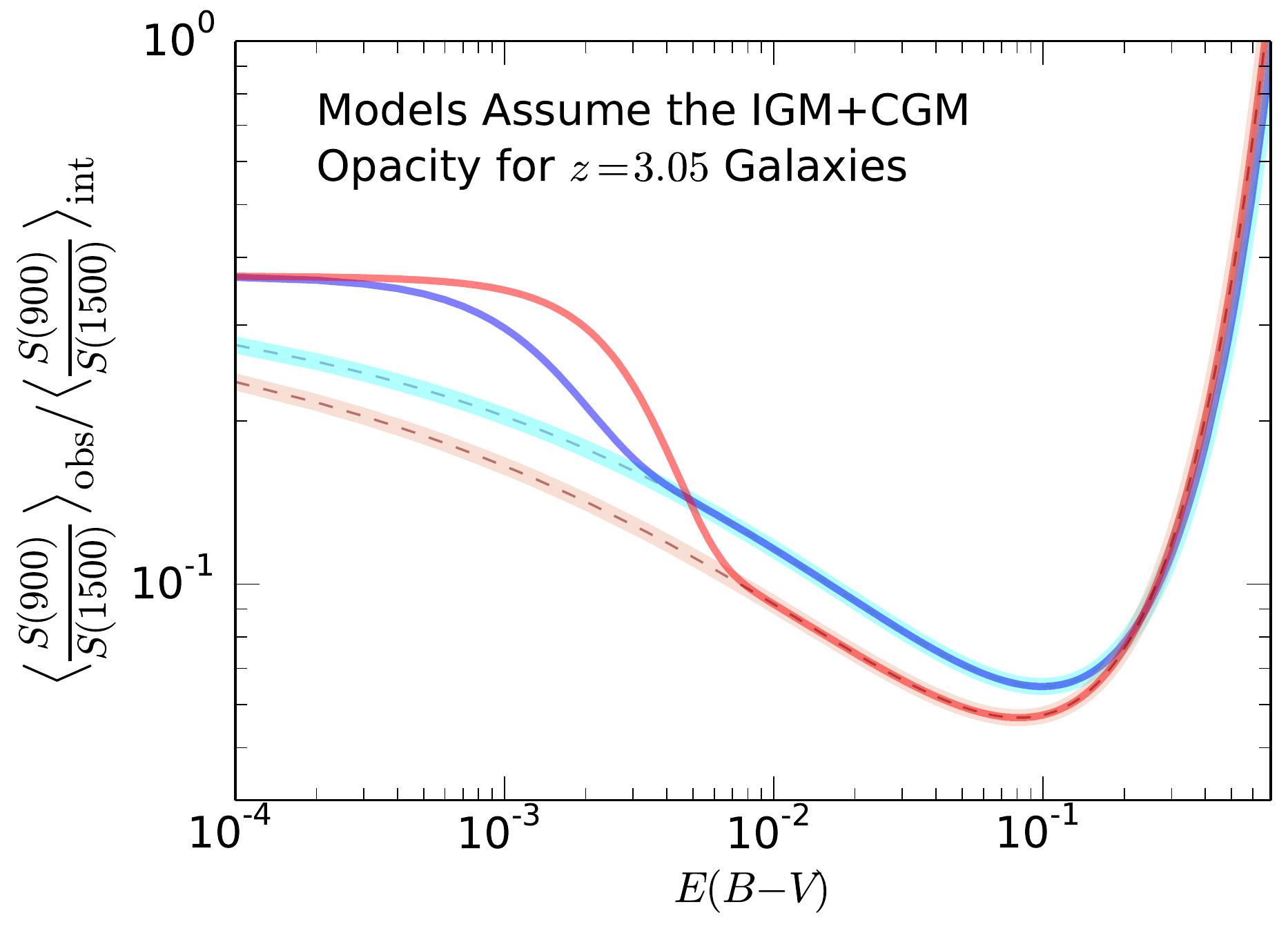}
\caption{Dependence of the observed 900-to-1500\,\AA\, flux density
  ratio (normalized by the intrinsic ratio) on continuum color excess,
  $\ebmv$, assuming the SMC ({\em blue} and {\em cyan} lines) or the
  Milky Way extinction curve ({\em red} and {\em light red} lines).
  The solid lines indicate the ``fiducial'' models where $\nhi$ varies
  with $\ebmv$ according to the relations listed in
  Table~\ref{tab:covfracfits}.  The thick shaded curves denote the
  ``extrapolated'' models which assume that the covering $\hi$ is
  optically-thick to ionizing photons for all values of $\ebmv$.  The
  dashed curves represent the ``fixed'' models, where $\nhi/\ebmv_{\rm
    los}$ is fixed and $\ebmv_{\rm los}$ varies with $\ebmv$ according
  to the relation listed in Table~\ref{tab:covfracfits}.  All curves
  assume the combined IGM and CGM opacities appropriate for the
  average line-of-sight toward $z\sim 3$ galaxies.}
\label{fig:models}
\end{figure}

Next, we use the observed relations listed in
Table~\ref{tab:covfracfits} to estimate $f_{\rm cov}(\hi)$,
$\ebmv_{\rm los}$ and $\nhi$, for different values of $\ebmv$.  Thus,
Equations~\ref{eq:twoterms}, \ref{eq:final1} and \ref{eq:final2} can
be written purely in terms of $\ebmv$, the line-of-sight extinction
and attenuation curves evaluated at the wavelengths of LyC and UV
photons, respectively, and the intrinsic LyC-to-UV flux density ratio.
To examine the possible range of models given the constraints imposed
by our data, we calculated an ``extrapolated'' model which assumes
that the neutral gas is optically-thick to ionizing photons at all
values of $\ebmv$, and that the uncovered portions of the galaxies are
effectively free of dust and gas.  In this case, the model is given
simply by Equation~\ref{eq:final2}.  In the ``fiducial'' model, we
assumed that the neutral gas column density varies with $\ebmv$
according to the relations listed in Table~\ref{tab:covfracfits}.
Finally, in the ``fixed'' model, we assumed that the ratio
$\nhi/\ebmv_{\rm los}$ is fixed, and that $\ebmv_{\rm los}$ scales
with $\ebmv$ according to the relations listed in
Table~\ref{tab:covfracfits}.  All three models were computed assuming
the combination of the updated \citet{reddy15} attenuation curve and
either the SMC or Milky Way extinction curves, and are shown in
Figure~\ref{fig:models}.  Other combinations of the stellar
attenuation and line-of-sight extinction curves yield quantitatively
similar results (see discussion below).
All models shown are computed by taking the UV luminosity density at
$1500$\,\AA\, and are cast in terms of the ratio of the observed and
intrinsic 900-to-1500\,\AA\, flux density ratios.

\subsubsection{Behavior of the Models}

In all of the cases considered here, the extrapolated and fixed models
are indistinguishable, because a fixed $\nhi/\ebmv_{\rm los}$ implies
optically thick $\hi$ over the entire range of $\ebmv=10^{-4}$ to
$10^{0}$ shown in Figure~\ref{fig:models}.  The fiducial model
deviates from the other two at $\ebmv \la 0.01$ where the neutral gas
becomes optically thin ($\nhi \la 5\times 10^{17}$\,cm$^{-2}$), thus
allowing the escape of additional LyC photons above the expectations
from the extrapolated model (i.e., the covered portion of the
galaxies' continua, as illustrated in Figure~\ref{fig:cartoon},
becomes optically thin).  For very blue $\ebmv$, $f_{\rm
  cov}(\hi)\rightarrow 0$ and the observed LyC-to-UV flux density
ratio simply converges to the intrinsic value modulated by the
combined IGM and CGM opacities.  For very red $\ebmv$, the ratio of
the observed and intrinsic LyC-to-UV flux density ratios increases to
values larger than unity.  This can be explained as follows.  LyC
photons escape the galaxies only through clear sightlines, and the
modeling assumes that the dust attenuation of such photons in
negligible along these sightlines.  Thus, the only way in which the
observed LyC flux density depends on the reddening is through the
relationship between $f_{\rm cov}(\hi)$ and $\ebmv$.  The absolute
escape fraction of LyC photons, i.e., $1-f_{\rm cov}(\hi)$, does not
decrease as rapidly as the absolute escape fraction of non-ionizing UV
photons---as the covering fraction asymptotes to unity for
$\ebmv\rightarrow \infty$---resulting in a steep rise of the model
prediction of $[S(900)_{\rm obs}/S(1500)_{\rm obs}]/[S(900)_{\rm
    int}/S(1500)_{\rm int}]$ for $\ebmv \ga 0.3$.

There is negligible variation in the model curve shapes at $\ebmv\la
0.1$ for different assumptions of the stellar attenuation curve.
Thus, in practice, one can simply assume a single model that is
invariant to assumptions of the adopted attenuation curve.  On the
other hand, the offset in the relation between $f_{\rm cov}(\hi)$ and
$\ebmv$ that arises from different assumptions of the line-of-sight
extinction curve (e.g., Figure~\ref{fig:covfrac}) results in a factor
of up to $\approx 1.5$ difference in the predicted $\langle
S(900)/S(1500)\rangle_{\rm obs}$ at a given $\ebmv$ for $\ebmv\la
0.10$.  This systematic difference may be partly attributed to the
lack of data for galaxies bluer than $\ebmv \simeq 0.1$ with which to
constrain the fit between $f_{\rm cov}(\hi)$ and $\ebmv$.  In the
following, we discuss how these model curves may be used to evaluate
the possible mechanisms of LyC escape and/or constrain the intrinsic
ionizing spectrum of high-redshift galaxies.

\subsubsection{Constraints on the Intrinsic LyC-to-UV Flux Density Ratio}
\label{sec:applications}

We preface our discussion by noting that deep far-UV spectra of the
kind used for our study would be useful to obtain over a broader
dynamic range of $\ebmv$, in particular focusing on the bluer galaxies
that are more analogous to typical galaxies at $z\ga 5$ and which may
have larger ionizing escape fractions.  Far-UV spectra may be used to
directly measure the covering fraction of neutral gas at $\ebmv \la
0.09$ using the technique described in Section~\ref{sec:specfitting},
and will enable more accurate estimates of the column densities and
line-of-sight reddening for such galaxies.  Nonetheless, in the
following, we illustrate how such a model may be used to discern the
mechanisms by which LyC photons escape from high-redshift galaxies.

The most direct application of our model is that it can be used to
predict the average relative escape fraction of ionizing photons for a
given stellar population model.  For example, if we assume $\langle
S(900)/S(1500)\rangle_{\rm int} = 1/6$, a value typical of that
adopted in previous studies (e.g., \citealt{siana07, nestor11}), then
a sample of galaxies at $z=3$ with $\langle \ebmv\rangle = 0.15$ would
be predicted to have an observed relative escape fraction of $\langle
S(900)/S(1500)\rangle_{\rm obs} \approx 0.011$.  The relative escape
fraction corrected for attenuation by the IGM and CGM is $0.031$, and
the absolute escape fraction is $\approx 5\%$.  We emphasize that the
model should only be applied to an ensemble of galaxies where the
stellar populations and different configurations of the gas and stars
from galaxy-to-galaxy, and the stochasticity in the Ly$\alpha$ forest
along different sightlines, can be averaged out.

A second and potentially powerful application of our model is
immediately obvious.  Namely, with direct measurements of the average
observed LyC-to-UV flux density ratio for an ensemble of galaxies with
some average continuum reddening, $\langle\ebmv\rangle$, one may
invert Equation~\ref{eq:final2} to calculate $\langle
S(900)/S(1500)\rangle_{\rm int}$.  The importance of direct empirical
constraints on the intrinsic LyC-to-UV flux density ratio is
underscored by the fact that the different flavors of stellar
population models (e.g., arising from different ages and metallicities
for a given star-formation history, and/or the treatment of stellar
rotation/binarity; \citealt{eldridge09, brott11, levesque12,
  leitherer14}) predict a factor of $\approx 5$ variation in the
intrinsic LyC-to-UV flux density ratios of $\simeq 1.3-6.4$ (e.g.,
\citealt{nestor13}).  The actual range in the predictions of the
stellar population models are larger than a factor of $\approx 5$ if
we also consider variations in the stellar initial mass function (IMF)
and star-formation history.  Clearly, independent information from the
rest-frame UV continuum spectra, rest-frame optical nebular emission
line spectra, and/or stellar population modeling may be used to narrow
the range of possible $\langle S({\rm LyC})/S({\rm UV})\rangle_{\rm int}$
\citep{steidel16}.  Nevertheless, there is substantial
utility in directly constraining the intrinsic LyC-to-UV flux density
ratio without having to extrapolate the spectra of massive stars to
wavelengths blueward of what is typically accessible from
observations.

To demonstrate how our model can be used to constrain the intrinsic
LyC-to-UV flux density ratio, we considered the recent measurement of
the relative escape fraction of galaxies in our sample by Steidel
et~al. (in prep).  For the subsample of galaxies with deep LyC
spectra---where the sample has been cleaned of foreground
contamination based on the ground-based spectroscopy (see
Section~\ref{sec:sample})---these authors measured the ratio of the
flux densities at $900$ and $1500$\,\AA, $\langle
S(900)/S(1500)\rangle_{\rm obs} = 0.019\pm 0.002$.  With this
measurement, we can invert Equation~\ref{eq:final2} to compute
$\langle S(900)/S(1500)\rangle_{\rm int}$.  To account fully for the
measurement and random errors that enter into
Equation~\ref{eq:final2}, we performed a Monte Carlo simulation by
perturbing each variable in the equation by its associated measurement
or random uncertainty.  As noted above, the mean combined IGM and CGM
attenuation at the mean redshift of the 121 galaxies of $\langle
z\rangle = 3.05$ is $\langle \exp[-\tau_{\rm IGM}({\rm LyC})] \times
\exp[-\tau_{\rm CGM}({\rm LyC})]\rangle = 0.370 \pm 0.030$.  The mean
reddening of these galaxies is $\langle \ebmv\rangle = 0.212\pm
0.003$, assuming the \citet{reddy15} dust attenuation curve.  For this
attenuation curve, $k({\rm 1500\,\AA}) = 8.687\pm0.200$ as calculated
in Paper~I.

\begin{figure}
\epsscale{1.15}
\plotone{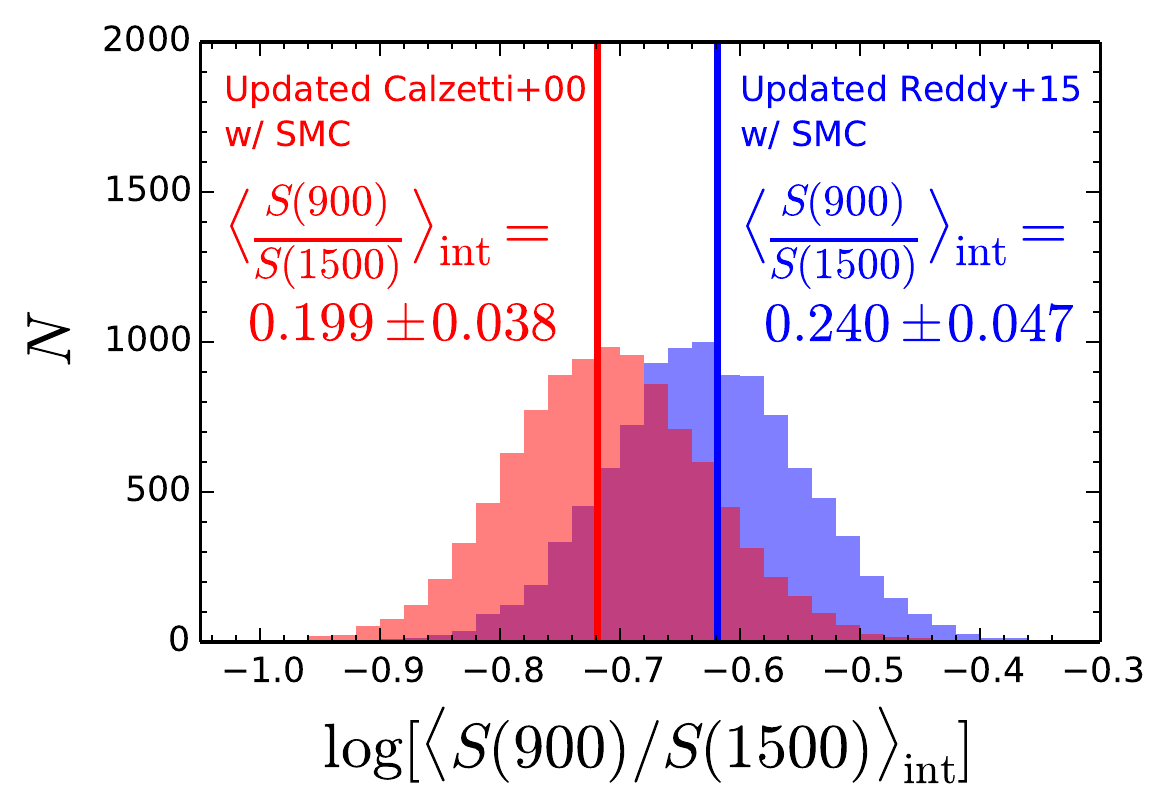}
\caption{Distribution of the intrinsic average $900$-to-$1500$\,\AA\,
  flux density ratios calculated using Equation~\ref{eq:final2} for
  10,000 trials.  In each trial, we perturbed the attenuation by the
  IGM$+$CGM, the mean reddening, the relationship between $f_{\rm
    cov}$ and $\ebmv$, the value of the attenuation curve at
  $1500$\,\AA\, ($k(1500)$), and the observed relative escape fraction
  of $\langle S(900)/S(1500)\rangle_{\rm obs} = 0.019\pm0.002$ from
  Steidel et~al. (in prep), each according to the random/measurement
  errors.  Shown are the distributions when we assume the updated
  \citet{reddy15} ({\em blue}) and \citet{calzetti00} ({\em red})
  attenuation curves for the stellar continuum, and an SMC extinction
  curve for the line-of-sight reddening.  Indicated are the means
  ({\em vertical lines}) and $1$\,$\sigma$ dispersions of the
  intrinsic $900$-to-$1500$\,\AA\, flux density ratios based on the
  10,000 trials.  The distributions obtained with the original
  versions of the \citet{reddy15} and \citet{calzetti00} attenuation
  curves for the reddening of the stellar continuum and/or by assuming
  a MW extinction curve for the line-of-sight reddening are
  essentially identical to the distributions shown here.}
\label{fig:inthist}
\end{figure}

We performed 10,000 simulations where, each time, we randomly
perturbed the IGM$+$CGM attenuation, the mean reddening, the dust
attenuation curve, and the observed relative escape fraction according
to normal distributions centered on the mean values with $\sigma$
equal to the associated uncertainties.  In addition, we randomly
perturbed $f_{\rm cov}(\hi)$ and $\ebmv$ for each composite according
to their respective uncertainties and fit Equation~\ref{eq:functional}
to these perturbed values.  We then calculated $\langle
S(900)/S(1500)\rangle_{\rm int}$ for each of the 10,000 trials.  The
resulting distributions of $\langle S(900)/S(1500)\rangle_{\rm int}$
are shown in Figure~\ref{fig:inthist}.  We find $\langle
S(900)/S(1500)\rangle_{\rm int} = 0.199\pm 0.038$ and $0.240\pm 0.047$
assuming the \citet{calzetti00} and \citet{reddy15} attenuation
curves, respectively (with an SMC line-of-sight extinction in both
cases).  The simulations indicate that we can use our methodology to
constrain the intrinsic 900-to-1500\,\AA\, flux density ratio to $\la
20\%$ {\em random} error for the sample of galaxies considered here.

%

In addition to this random error, there are several sources of
systematic error in our estimate of $\langle
S(900)/S(1500)\rangle_{\rm int}$, including the choice of the
line-of-sight extinction curve (e.g., SMC or MW) and the choice of the
stellar reddening curve (e.g., \citealt{reddy15} or
\citet{calzetti00}).  The magnitude of the bias in $\langle
S(900)/S(1500)\rangle_{\rm int}$ that stems from these effects is $\la
2\%$ and $\approx 25\%$, respectively.  Additional systematic
uncertainty may arise from the choice of IGM prescription.  For
example, \citet{inoue14} find a mean IGM transmission of $\approx 0.5$
for $900$\,\AA\, photons emitted from a source at $z=3$, a value at
the high end of published measurements.  On the other hand, the {\em
  combined} IGM+CGM opacity predicted by \citet{rudie13} is $0.37$.
This suggests a systematic uncertainty associated with the IGM
prescription of $\approx 30\%$.

The largest source of systematic uncertainty relates to our inability
to fully resolve the $\hi$ absorption.  Specifically, our estimate of
$f_{\rm cov}(\hi)$ may be a lower limit as a result of several
effects, some of which are discussed in
Section~\ref{sec:evidencefornonunity}.  First, the relatively low
spectral resolution of our data may mask the presence of narrow
absorption line systems with a higher covering fraction than that
deduced from the spectra.  Second, if the different velocity
components of the optically-thick $\hi$ are not covering the same
lines-of-sight, then the integrated covering fraction may be larger
than the value found for the bulk of the absorbing gas.  Third, we are
unable to discern whether non-negligible columns of $\hi$ and dust
exist in regions that we would otherwise refer to as ``clear''
sightlines.  Fourth, the ionized gas may not be completely bounded by
$\hi$ (i.e., the ``average'' configuration of the $\hi$ and \ion{H}{2}
may be intermediate between the radiation- and density-bounded
scenarios).  In this latter case, attenuation by dust in the ionized
gas may deplete LyC photons before they escape the galaxy.  For these
reasons, the {\em effective} covering fraction, $f_{\rm cov}^{\rm
  eff}(\hi)$, may be larger than the value derived from the composite
spectrum and hence the calculated value of $\langle
S(900)/S(1500)\rangle_{\rm int}$ may be a lower limit.

\begin{figure}
\epsscale{1.10}
\plotone{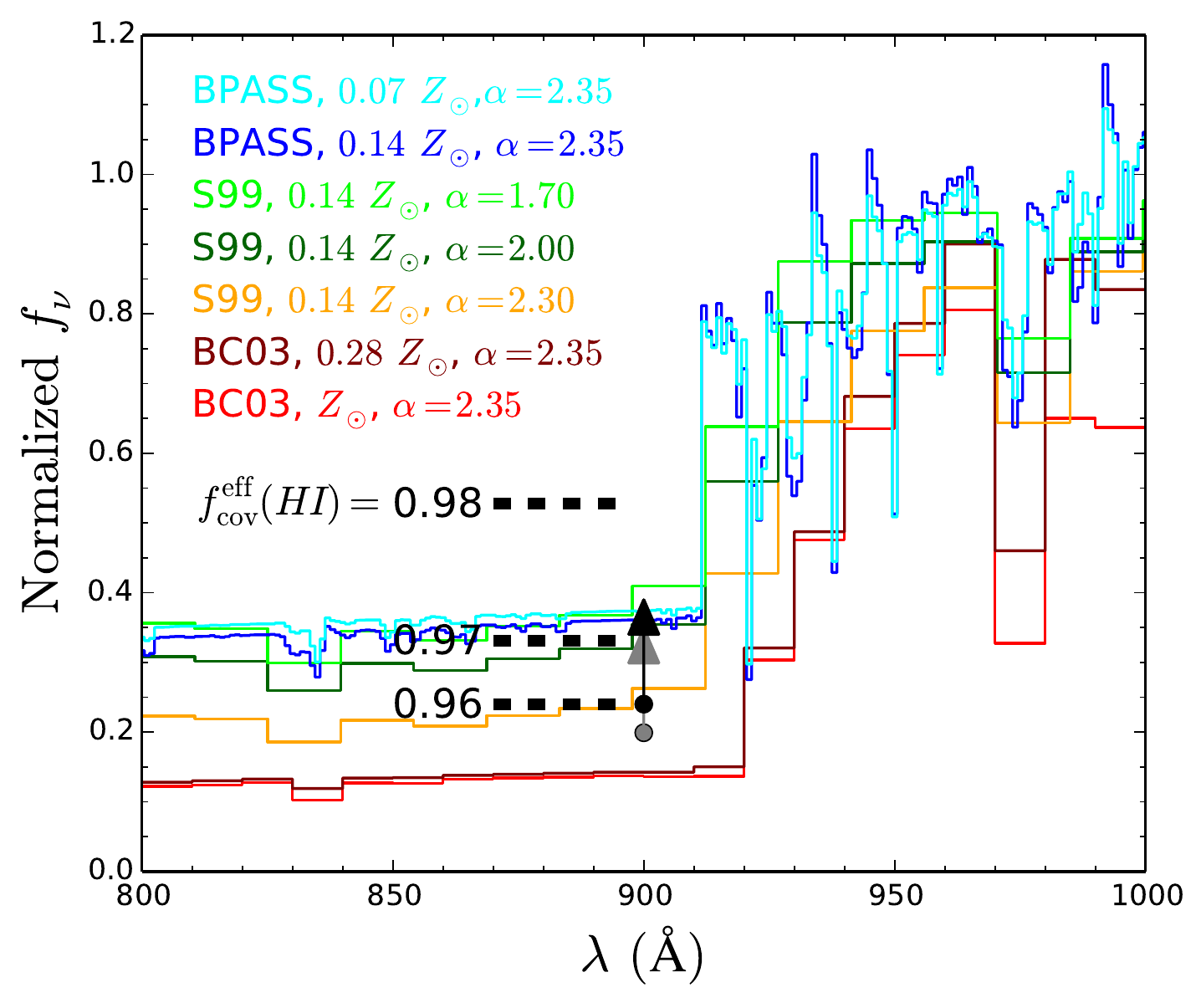}
\caption{Comparison of different stellar population models, including
  those of \citet{bruzual03}, the latest version of Starburst99
  \citep{leitherer14}, and the latest version of the BPASS models
  \citep{stanway16}, at extreme- to far-UV wavelengths, for different
  metallicities and power law indices ($\alpha$) of the IMF.  All of
  the models are normalized to have a unity flux density at
  $1500$\,\AA.  The limits in the intrinsic 900-to-1500\,\AA\, flux
  density ratio, $\langle S(900)/S(1500)\rangle_{\rm int}$,
  as inferred from our spectra for an SMC line-of-sight extinction and
  the updated \citet{reddy15} and \citet{calzetti00} stellar
  attenuation curves are indicated by the black and grey arrows,
  respectively.  The covering fraction deduced from the depth of the
  $\hi$ lines in the composite spectrum of all galaxies in our sample
  is $f_{\rm cov}(\hi) = 0.96$.  If the ``effective'' covering
  fraction is $f_{\rm cov}^{\rm eff} = 0.97$ and $0.98$, then the
  limit in $\langle S(900)/S(1500)\rangle_{\rm int}$ rises
  to $0.33$, and $0.53$, respectively, as indicated by the horizontal
  dashed lines.}
\label{fig:stelpop}
\end{figure}

To further evaluate the range of possible $\langle
S(900)/S(1500)\rangle_{\rm int}$ allowed by the data, we compared our
estimate with the predictions of several stellar population synthesis
models (Figure~\ref{fig:stelpop}).  We considered \citet{bruzual03}
$0.28$\,$Z_\odot$ and $Z_\odot$ models with a \citet{salpeter55} IMF
(power law index of $\alpha=2.35$) over the mass range $0.1\le
M_{\ast} \le 100$\,$M_\odot$; and the most recent version of the
Starburst99 (S99) models which adopt weaker stellar winds relative to
the the previous version \citep{leitherer14}.  We generated S99 models
using the \citet{kroupa01} IMF with different power law slopes of
$2.30$, $2.00$, and $1.70$, over the mass range $0.5\le M_{\ast} \le
100$\,$M_\odot$, and with a metallicity of $0.14$\,$Z_\odot$.  Lastly,
we consider the ``Binary Population and Spectral Synthesis'' (BPASS;
\citealt{stanway16}) models that include binary evolution, with a
power law slope of the IMF of $2.35$ over the mass range $0.5\le
M_{\ast} \le 300$\,$M_\odot$, with metallicities of $0.07$ and
$0.14$\,$Z_\odot$.  All metallicities assume the solar abundances
measured in \citet{asplund09}.

The effect of binary evolution (as in the BPASS models) is to allow
efficient mass transfer from a massive star to its (secondary)
companion, causing the latter to spin up quickly and, for a low
stellar metallicity where the stellar wind is sufficiently weak,
retain its angular momentum \citep{cantiello07}.  Consequently, the
secondary becomes fully mixed and undergoes quasi-homogeneous
evolution \citep{yoon05, brott11, eldridge12} whereby the main
sequence lifetime and effective temperature are increased.  In such
models, the ionizing photon flux per non-ionizing UV photon flux may
increase by up to $60\%$ relative to single star evolution (e.g.,
\citealt{stanway16}; see also \citealt{topping15}).  A similar effect
can be achieved by decreasing the power law index of the IMF, which
results in a larger mass fraction in more massive, hotter stars.

Figure~\ref{fig:stelpop} shows the limit on $\langle
S(900)/S(1500)\rangle_{\rm int}$ based on $f_{\rm cov}(\hi) \approx
0.96$ derived from the composite spectrum of all 933 galaxies in our
sample.  Our inference of the gas covering fraction favors an
intrinsic LyC-to-UV flux density ratio that is larger than the value
predicted from the \citet{bruzual03} stellar population models, and
one that is consistent with those obtained with the newer versions of
the S99 models that adopt weaker massive star stellar winds, a low
stellar metallicity, and/or a flatter slope of the IMF; or those
models that include the effects of binary evolution.  For the reasons
stated above, the covering fractions derived here are likely lower
limits, and therefore it is instructive to gauge the degree to which
the covering fraction may be higher and still be consistent with the
predictions of stellar population synthesis models.

In particular, if the effective covering fraction is $f_{\rm cov}^{\rm
  eff}(\hi) \approx 0.97$ or $0.98$, instead of $0.96$, then the
inferred $\langle S(900)/S(1500)\rangle_{\rm int}$ increases from
$0.24$ to $0.31$ and $0.51$, respectively.  Note that the latter is
$\approx 50\%$ larger than the value predicted by the low-metallicity
BPASS models and/or the low-metallicity and shallow-$\alpha$ IMF S99
models.  Thus it is apparent, though ironic, that it is those galaxies
with the largest covering fractions (as is the case for the average
galaxy at $z\sim 3$) which provide the most sensitive constraints on
the intrinsic LyC-to-UV flux density ratio, since even small increases
in $f_{\rm cov}(\hi)$ will lead to very large changes in the inferred
$\langle S(900)/S(1500)\rangle_{\rm int}$ (Figure~\ref{fig:stelpop}).
\citet{steidel16} suggest that stellar population models that
include the effects of binary evolution are required to simultaneously
reproduce the rest-frame far-UV continuum, stellar, and nebular
features, and the rest-frame optical nebular emission line strengths
measured for galaxies at $z\sim 2$.  If we adopt such models as the
most appropriate, then it implies that our estimate of $f_{\rm
  cov}(\hi)$ for typical galaxies at $z\sim 3$ is not biased low by
more than $\approx 1\%$.  Stated another way, the combined effects of
additional $\hi$ absorption and dust attenuation cannot deplete more
than $25\%$ of the photons exiting through what we would otherwise
refer to as ``clear'' sightlines based on the depths of the $\hi$
lines in the composite spectra.

{\em This analysis underscores the fact that if the simple geometry of
  Figure~\ref{fig:cartoon} is valid on average for an ensemble of
  galaxies, then deep UV spectroscopy for large samples of typical
  star-forming galaxies at high redshift should enable tight
  constraints on the intrinsic LyC-to-UV flux density ratio.}  Hence,
our framework provides an alternative and powerful independent method
of constraining the intrinsic LyC-to-UV flux density ratio, which is
currently one of the most uncertain parameters entering into the
calculation of the absolute escape fraction of ionizing photons in
high-redshift galaxies.

\begin{figure}
\epsscale{1.15}
\plotone{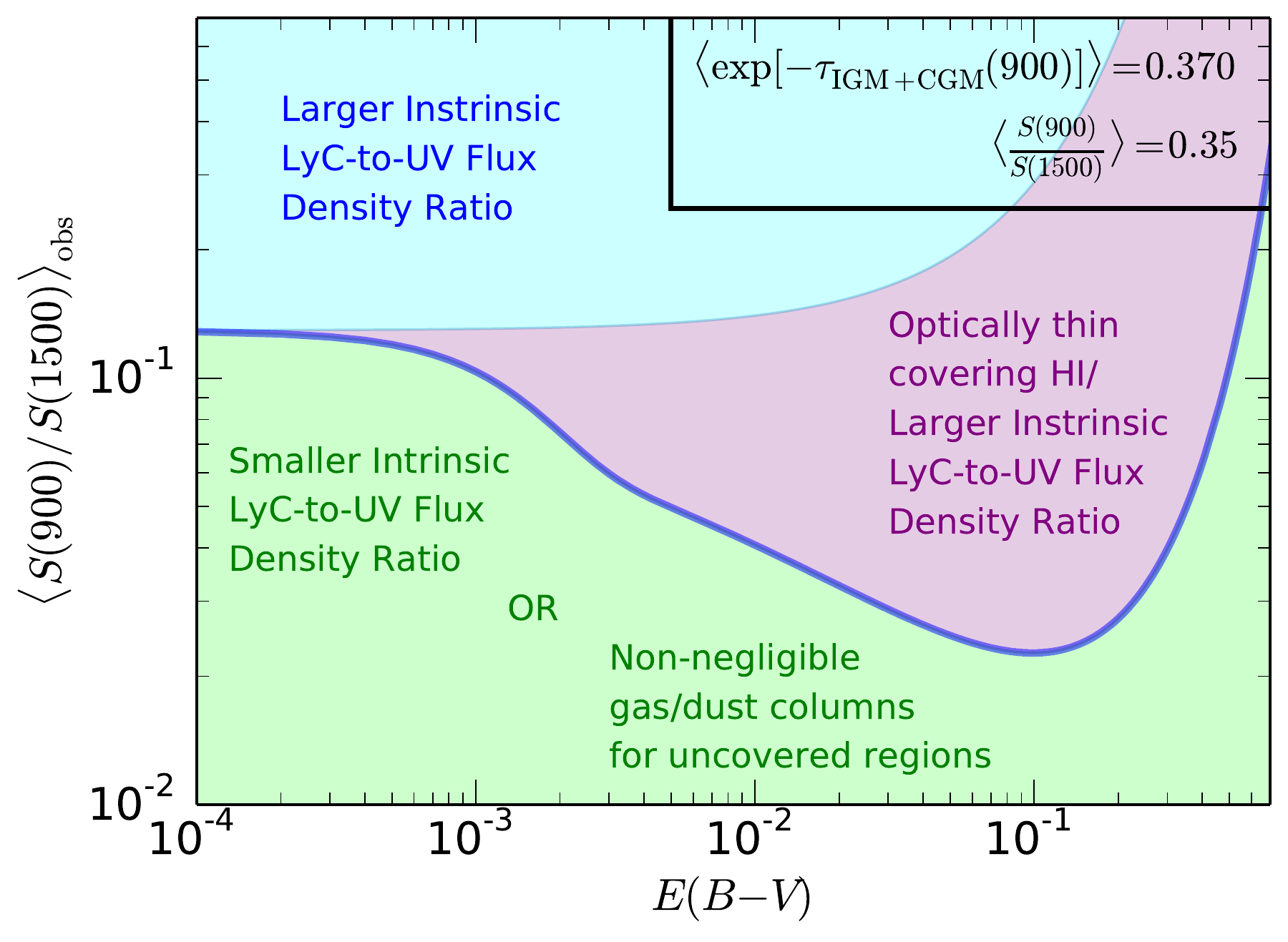}
\caption{Same as Figure~\ref{fig:models}, where only the fiducial
  model for the SMC extinction curve ({\em blue} line) is shown for
  simplicity, and we have assumed $\langle S(900)/S(1500)\rangle_{\rm
    int} = 0.35$.  As before, an IGM+CGM opacity appropriate for
  galaxies at $z=3.05$ is assumed.  A measured $\langle S({\rm
    LyC})/S({\rm UV})\rangle_{\rm obs}$ and $\langle
  \ebmv\rangle $ in the {\em green} shaded for an ensemble of galaxies
  implies either $\langle S(900)/S(1500)\rangle_{\rm int}< 0.35$, or
  that the ``uncovered'' portions of the galaxies' spectra have a
  non-negligible column density of gas and/or dust.  Galaxies with
  average values that place them in the {\em purple} region may have a
  larger intrinsic 900-to-1500\,\AA\, flux density ratio {\em or}
  outflowing neutral gas that is optically thin (allowing for the
  leakage of additional LyC photons).  The boundary between the {\em
    cyan} and {\em purple} regions is defined by setting $f(900)_{\rm
    abs} = 1$, so that if the average values place the galaxies in the
  {\em cyan} region, then such galaxies {\em must} have a larger
  intrinsic 900-to-1500\,\AA\, flux density ratio than the one assumed
  here.}
\label{fig:devgeo}
\end{figure}

\subsubsection{Deviations from the Assumed Geometry}
\label{sec:devgeo}

The geometry depicted in Figure~\ref{fig:cartoon} may not apply for
all galaxies, and in reality more complicated configurations of the
dust, gas, and stars may be expected.  In this context, the model
described here provides a useful benchmark to investigate changes in
geometry and hence alternate pathways for the escape of ionizing
radiation.  To illustrate this, we show in Figure~\ref{fig:devgeo}
$\langle S(900)/S(1500)\rangle_{\rm obs}$ as a function of $\ebmv$
assuming the fiducial model for the SMC extinction curve.  We have
assumed $\langle S(900)/S(1500)\rangle_{\rm int}=0.35$, a value
typical of the BPASS models and the S99 models with a flat IMF slope,
and IGM and CGM opacities relevant for galaxies at z=3.05.  Direct
measurements of $\langle S(900)/S(1500)\rangle_{\rm obs}$ and
$\langle\ebmv\rangle$ for an {\em ensemble} of galaxies may be used to
ascertain the mechanisms by which LyC photons escape galaxies.  For
example, if these average values lie in the {\em green} region, then
it suggests that, at least on average, those galaxies have either a
smaller intrinsic 900-to-1500\,\AA\, flux density ratio than the one
we have adopted here, or that the ``uncovered'' portions of the
continuum are not completely free of dust and/or gas.  Similarly,
objects whose average values place them in the {\em purple} region are
ones that may have a larger intrinsic 900-to-1500\,\AA\, flux density
ratio or optically-thin neutral gas that allows the escape of
additional LyC photons.  Finally, galaxies whose average values lie in
the {\em cyan} region {\em must} have a larger intrinsic
900-to-1500\,\AA\, flux density ratio on average than the one assumed
here.  Further discrimination of the mechanisms by which LyC photons
escape galaxies, or of the value of the intrinsic LyC-to-UV flux
density ratio, can be accomplished with deep spectroscopy, such as
that presented here.  From this spectroscopy, we have found that
typical $z\sim 3$ galaxies can be characterized by optically-thick
outflowing neutral $\hi$, a finding that already narrows the range of
possible avenues by which ionizing radiation can escape these
galaxies.

\section{CONCLUSIONS}

In this paper, we have modeled the rest-frame far-UV composite spectra
of 933 $\rs < 25.5$ LBGs at $z\sim 3$, 121 of which have very deep
spectroscopic covering at $\lambda = 850-1300$\,\AA, in order to
investigate the relationship between UV reddening, neutral gas
covering fraction, and the ionizing photon escape fraction.  This
analysis is facilitated by our large sample of galaxies which allows
us to average over stochastic variations in the Ly$\alpha$ forest
along different sightlines to $z\sim 3$ galaxies, and thus securely
analyze the $\hi$ absorption lines in these galaxies.  Our main conclusions
are as follows.

\begin{itemize}

\item{The composite spectra of $z\sim 3$ galaxies show evidence for a
  partial covering of optically-thick outflowing neutral hydrogen.
  Specifically, the gas is (a) blueshifted by $\approx
  300$\,km\,s$^{-1}$; (b) exhibits damping wings in the Lyman series
  lines, including Ly$\alpha$, signifying high-column density,
  $\log[\nhi/{\rm cm^{-2}}] \ga 20.3$; and (c) Lyman series lines
  including Ly$\beta$, Ly$\gamma$, and Ly$\delta$ that do not reach
  zero flux density at the line cores, implying the presence of
  unabsorbed stellar continuum.  This result is robust against
  spectral resolution and redshift uncertainties, and is based upon a
  careful isolation of foreground contaminants.}

\item{To investigate trends between reddening and gas covering
  fraction, we divided our sample into bins of $\ebmv$ (measured from
  the stellar continuum) and constructed composites for galaxies in
  each of these bins.  We modeled the composite spectra, allowing for
  a non-unity covering fraction of $\hi$ and dust---a configuration
  that arises from an increase in ISM porosity due to stellar feedback
  processes; and including both stellar and nebular continuum
  emission, as well as unresolved absorption from the $\molh$
  Lyman-Werner bands.  The derived $\hi$ covering fractions vary from
  $92$ to $97\%$, over a range of continuum reddening from $\ebmv =
  0.098$ to $0.292$.  The modeling results imply that galaxies with
  redder UV continua have a larger covering fraction of $\hi$ and
  dust, with a higher line-of-sight dust column density.}

\item{The absolute escape fraction of Ly$\alpha$ emission, as measured
  through a spectroscopic slit, correlates inversely with neutral gas
  covering fraction.  This suggests that the Ly$\alpha$ photons escape
  preferentially through sightlines that have been cleared of gas and
  dust.  Separately, the neutral gas covering fractions deduced from
  saturated low-ionization interstellar absorption lines, such as
  \ion{Si}{2}\,$\lambda\lambda 1260, 1527$\,\AA, correlate with, but
  systematically underpredict by a factor of $\simeq 1.5$, those
  derived from the $\hi$ lines.  Thus, inferences of gas covering
  fractions based on low-ionization interstellar absorption lines must
  be viewed with caution.  Moreover, this result suggests that a
  significant fraction of outflowing $\hi$ may be metal-poor.  The
  fact that the ``covering fractions'' inferred from the
  low-ionization interstellar absorption lines appear to rise rapidly
  with a small increase in $f_{\rm cov}(\hi)$, together with our
  finding that the line-of-sight extinction increases in tandem with
  the reddening of the stellar continuum, suggests that the outflowing
  material becomes progressively enriched with metals for galaxies
  with larger $f_{\rm cov}(\hi)$ and redder UV continua.}

\item{We discuss our results in the context of the escape of ionizing
  radiation at high redshift.  The partial covering fraction of
  optically-thick $\hi$ implies that the depletion of ionizing photons
  is dominated by photoelectric absorption, rather than dust
  attenuation, and that ionizing photons escape typical galaxies at
  $z\sim 3$ through clear sightlines in the ISM.}

\item{Using a physically-motivated functional form for the
  relationship between continuum reddening and neutral gas covering
  fraction, we establish a relationship between the absolute and
  relative escape fraction of ionizing radiation and continuum
  reddening.  Our model may be used to predict the average escape
  fractions of ionizing photons for an ensemble of galaxies with some
  average $\ebmv$ and an assumed intrinsic LyC-to-UV flux density
  ratio.  Alternatively, direct measurements of the ionizing escape
  fraction may be used to place constraints on the intrinsic LyC-to-UV
  flux density ratio.  Based on a recent measurement of $\langle
  S(900)/S(1500)\rangle_{\rm obs}$ for the subsample of galaxies with
  deep far-UV spectra analyzed here, we infer $\langle
  S(900)/S(1500)\rangle_{\rm int} \ga 0.20$.  This lower limit is
  larger than the prediction from the \citet{bruzual03} models, and
  generally favors stellar population models that include weaker
  stellar winds, a flatter slope of the IMF, and/or binary evolution.}

\end{itemize}

We have demonstrated how our model may be used to provide powerful
constraints on the intrinsic ionizing spectrum of high-redshift
galaxies and, additionally, how galaxies with large covering fractions
and associated large reddening, may be used as sensitive probes of the
intrinsic LyC-to-UV flux density ratio.  Aside from providing
constraints on the production efficiency of ionizing photons, we have
discussed how our model may also be used to assess the pathways by
which such photons escape galaxies.

Our analysis points to several future avenues of investigation.
Coupling deep rest-frame UV spectroscopy of high-redshift galaxies
(such as that presented here) with multi-band {\em HST} imaging will
allow us to construct the cleanest possible sample of
galaxies---uncontaminated by foreground interlopers based on high
spatial resolution multi-component photometric redshift analysis
\citep{mostardi15} and the identification of spectroscopic
blends---with which to measure the ionizing escape fraction.  The high
spatial resolution provided by {\em HST} has other potential benefits
in terms of clarifying the escape of ionizing radiation.
Specifically, a consequence of a partial covering fraction of $\hi$ is
an inhomogeneous spatial distribution of LyC emission relative to the
non-ionizing UV continuum.  While resolving individual sightlines in
these high-redshift galaxies is far beyond {\em HST}'s capabilities, a
``clumpy'' distribution of LyC emission that is resolved with {\em
  HST} to be disjoint from the morphology of the non-ionizing UV
continuum may still indicate the signature of a porous ISM where LyC
photons only escape through clear lines-of-sight.

More precise measurements of, or limits on, the neutral gas covering
fractions may be achieved with higher resolution spectroscopy for
large samples of galaxies, possible with the next-generation of
instruments and $\ga 30$\,m telescopes.  In the more immediate future,
rest-frame optical nebular emission line spectroscopy will enable
systemic redshift measurements that can be used to analyze the $\hi$
line profiles from composite spectra in a way that is less affected by
redshift uncertainties.  The combination of rest-frame optical nebular
spectroscopy, rest-frame UV continuum spectroscopy, and our model for
the escape of ionizing radiation together will make it possible to
construct a consistent picture of the massive and ionizing stellar
populations in high-redshift galaxies.  Lastly, the models presented
here will benefit greatly from future deep spectroscopy of bluer
galaxies and/or those detected with strong Ly$\alpha$ emission to
extend the dynamic range of continuum reddening probed in our study.
With such spectroscopy, we will be in a position to evaluate the
column densities, line-of-sight reddening, and covering fractions for
the types of galaxies that are likely to contribute significantly to
the budget of ionizing photons at high redshift.  The framework
presented in this paper will be central to the interpretation of the
ionizing escape fractions of galaxies at the reionization epoch, where
the prohibitively thick IGM precludes the detailed far-UV measurements
presented here.

\acknowledgements

NAR is supported by an Alfred P. Sloan Research Fellowship, and
acknowledges the visitors program at the Institute of Astronomy in
Cambridge, UK, where part of this research was conducted.  CCS
acknowledges NSF grants AST-0908805 and AST-1313472.  MB acknowledges
support of the Serbian MESTD through grant ON176021.  NAR acknowledges
Gwen Rudie for useful discussions.  We are grateful to the anonymous
referee whose comments led to significant improvements in the clarity,
content, and presentation of the analysis and results.  We wish to
extend special thanks to those of Hawaiian ancestry on whose sacred
mountain we are privileged to be guests.  Without their generous
hospitality, the observations presented herein would not have been
possible.


\begin{thebibliography}{102}
\expandafter\ifx\csname natexlab\endcsname\relax\def\natexlab#1{#1}\fi

\bibitem[{{Alexandroff} {et~al.}(2015){Alexandroff}, {Heckman}, {Borthakur},
  {Overzier}, \& {Leitherer}}]{alexandroff15}
{Alexandroff}, R.~M., {Heckman}, T.~M., {Borthakur}, S., {Overzier}, R., \&
  {Leitherer}, C. 2015, \apj, 810, 104

\bibitem[{{Armus} {et~al.}(1995){Armus}, {Heckman}, {Weaver}, \&
  {Lehnert}}]{armus95}
{Armus}, L., {Heckman}, T.~M., {Weaver}, K.~A., \& {Lehnert}, M.~D. 1995, \apj,
  445, 666

\bibitem[{{Asplund} {et~al.}(2009){Asplund}, {Grevesse}, {Sauval}, \&
  {Scott}}]{asplund09}
{Asplund}, M., {Grevesse}, N., {Sauval}, A.~J., \& {Scott}, P. 2009, \araa, 47,
  481

\bibitem[{{Bagetakos} {et~al.}(2011){Bagetakos}, {Brinks}, {Walter},
    {de Blok}, {Usero}, {Leroy}, {Rich}, \& {Kennicutt}}]{bagetakos11}
  {Bagetakos}, I., {Brinks}, E., {Walter}, F., et al. 2011, \aj, 141,
  23

\bibitem[{{Bolton} \& {Haehnelt}(2007)}]{bolton07}
{Bolton}, J.~S. \& {Haehnelt}, M.~G. 2007, \mnras, 382, 325

\bibitem[{{Bolton} {et~al.}(2005){Bolton}, {Haehnelt}, {Viel}, \&
  {Springel}}]{bolton05}
{Bolton}, J.~S., {Haehnelt}, M.~G., {Viel}, M., \& {Springel}, V. 2005, \mnras,
  357, 1178

\bibitem[{{Borthakur} {et~al.}(2014){Borthakur}, {Momjian}, {Heckman},
    {York}, {Bowen}, {Yun}, \& {Tripp}}]{borthakur14} {Borthakur}, S.,
  {Momjian}, E., {Heckman}, T.~M., et al. 2014, \apj, 795, 98

\bibitem[{{Bouwens} {et~al.}(2015){Bouwens}, {Illingworth}, {Oesch},
    {Trenti}, {Labb{\'e}}, {Bradley}, {Carollo}, {van Dokkum},
    {Gonzalez}, {Holwerda}, {Franx}, {Spitler}, {Smit}, \&
    {Magee}}]{bouwens15} {Bouwens}, R.~J., {Illingworth}, G.~D.,
  {Oesch}, P.~A., et~al. 2015, \apj, 803, 34

\bibitem[{{Bouwens} {et~al.}(2004){Bouwens}, {Thompson},
    {Illingworth}, {Franx}, {van Dokkum}, {Fan}, {Dickinson},
    {Eisenstein}, \& {Rieke}}]{bouwens04} {Bouwens}, R.~J.,
  {Thompson}, R.~I., {Illingworth}, G.~D., et al. 2004, \apjl, 616,
  L79

\bibitem[{{Bridge} {et~al.}(2010){Bridge}, {Teplitz}, {Siana},
    {Scarlata}, {Conselice}, {Ferguson}, {Brown}, {Salvato}, {Rudie},
    {de Mello}, {Colbert}, {Gardner}, {Giavalisco}, \&
    {Armus}}]{bridge10} {Bridge}, C.~R., {Teplitz}, H.~I., {Siana},
  B., et al. 2010, \apj, 720, 465

\bibitem[{{Brott} {et~al.}(2011){Brott}, {de Mink}, {Cantiello}, {Langer}, {de
  Koter}, {Evans}, {Hunter}, {Trundle}, \& {Vink}}]{brott11}
{Brott}, I., {de Mink}, S.~E., {Cantiello}, M., et al. 2011, \aap,
  530, A115

\bibitem[{{Bruzual} \& {Charlot}(2003)}]{bruzual03}
{Bruzual}, G. \& {Charlot}, S. 2003, \mnras, 344, 1000

\bibitem[{{Bunker} {et~al.}(2010){Bunker}, {Wilkins}, {Ellis},
    {Stark}, {Lorenzoni}, {Chiu}, {Lacy}, {Jarvis}, \&
    {Hickey}}]{bunker10} {Bunker}, A.~J., {Wilkins}, S., {Ellis},
  R.~S., et al. 2010, \mnras, 409, 855

\bibitem[{{Calzetti} {et~al.}(2000){Calzetti}, {Armus}, {Bohlin},
    {Kinney}, {Koornneef}, \& {Storchi-Bergmann}}]{calzetti00}
  {Calzetti}, D., {Armus}, L., {Bohlin}, R.~C., et al. 2000, \apj,
  533, 682

\bibitem[{{Cantiello} {et~al.}(2007){Cantiello}, {Yoon}, {Langer}, \&
  {Livio}}]{cantiello07}
{Cantiello}, M., {Yoon}, S.-C., {Langer}, N., \& {Livio}, M. 2007, \aap, 465,
  L29

\bibitem[{{Clarke} \& {Oey}(2002)}]{clarke02}
{Clarke}, C. \& {Oey}, M.~S. 2002, \mnras, 337, 1299

\bibitem[{{Cowie} {et~al.}(2009){Cowie}, {Barger}, \& {Trouille}}]{cowie09}
{Cowie}, L.~L., {Barger}, A.~J., \& {Trouille}, L. 2009, \apj, 692, 1476

\bibitem[{{de Barros} {et~al.}(2016){de Barros}, {Vanzella},
    {Amor{\'{\i}}n}, {Castellano}, {Siana}, {Grazian}, {Suh},
    {Balestra}, {Vignali}, {Verhamme}, {Zamorani}, {Mignoli},
    {Hasinger}, {Comastri}, {Pentericci}, {P{\'e}rez-Montero},
    {Fontana}, {Giavalisco}, \& {Gilli}}]{debarros16} {de Barros}, S.,
  {Vanzella}, E., {Amor{\'{\i}}n}, et al. 2016, \aap, 585, 51

\bibitem[{{Deharveng} {et~al.}(2001){Deharveng}, {Buat}, {Le Brun},
    {Milliard}, {Kunth}, {Shull}, \& {Gry}}]{deharveng01} {Deharveng},
  J.-M., {Buat}, V., {Le Brun}, V., et al. 2001, \aap, 375, 805

\bibitem[{{Duncan} \& {Conselice}(2015)}]{duncan15}
{Duncan}, K. \& {Conselice}, C.~J. 2015, \mnras, 451, 2030

\bibitem[{{Eldridge} \& {Stanway}(2009)}]{eldridge09}
{Eldridge}, J.~J. \& {Stanway}, E.~R. 2009, \mnras, 400, 1019

\bibitem[{{Eldridge} \& {Stanway}(2012)}]{eldridge12}
---. 2012, \mnras, 419, 479

\bibitem[{{Ellis} {et~al.}(2013){Ellis}, {McLure}, {Dunlop}, {Robertson},
  {Ono}, {Schenker}, {Koekemoer}, {Bowler}, {Ouchi}, {Rogers}, {Curtis-Lake},
  {Schneider}, {Charlot}, {Stark}, {Furlanetto}, \& {Cirasuolo}}]{ellis13}
{Ellis}, R.~S., {McLure}, R.~J., {Dunlop}, J.~S., et al. 2013, \apjl, 763, L7

\bibitem[{{Fan} {et~al.}(2001){Fan}, {Narayanan}, {Lupton}, {Strauss}, {Knapp},
  {Becker}, {White}, {Pentericci}, {Leggett}, {Haiman}, {Gunn}, {Ivezi{\'c}},
  {Schneider}, {Anderson}, {Brinkmann}, {Bahcall}, {Connolly}, {Csabai}, {Doi},
  {Fukugita}, {Geballe}, {Grebel}, {Harbeck}, {Hennessy}, {Lamb}, {Miknaitis},
  {Munn}, {Nichol}, {Okamura}, {Pier}, {Prada}, {Richards}, {Szalay}, \&
  {York}}]{fan01}
{Fan}, X., {Narayanan}, V.~K., {Lupton}, R.~H., et al. 2001, \aj, 122, 2833

\bibitem[{{Finkelstein} {et~al.}(2012){Finkelstein}, {Papovich}, {Ryan},
  {Pawlik}, {Dickinson}, {Ferguson}, {Finlator}, {Koekemoer}, {Giavalisco},
  {Cooray}, {Dunlop}, {Faber}, {Grogin}, {Kocevski}, \&
  {Newman}}]{finkelstein12b}
{Finkelstein}, S.~L., {Papovich}, C., {Ryan}, R.~E., et al. 2012, \apj, 758, 93

\bibitem[{{Fontanot} {et~al.}(2012){Fontanot}, {Cristiani}, \&
  {Vanzella}}]{fontanot12}
{Fontanot}, F., {Cristiani}, S., \& {Vanzella}, E. 2012, \mnras, 425, 1413

\bibitem[{{Giallongo} {et~al.}(2015){Giallongo}, {Grazian}, {Fiore}, {Fontana},
  {Pentericci}, {Vanzella}, {Dickinson}, {Kocevski}, {Castellano}, {Cristiani},
  {Ferguson}, {Finkelstein}, {Grogin}, {Hathi}, {Koekemoer}, {Newman}, \&
  {Salvato}}]{giallongo15}
{Giallongo}, E., {Grazian}, A., {Fiore}, F., et al. 2015, \aap, 578,
  A83

\bibitem[{{Glikman} {et~al.}(2011){Glikman}, {Djorgovski}, {Stern}, {Dey},
  {Jannuzi}, \& {Lee}}]{glikman11}
{Glikman}, E., {Djorgovski}, S.~G., {Stern}, D., et al. 2011, \apjl, 728, L26

\bibitem[{{Gnedin} {et~al.}(2008){Gnedin}, {Kravtsov}, \& {Chen}}]{gnedin08}
{Gnedin}, N.~Y., {Kravtsov}, A.~V., \& {Chen}, H.-W. 2008, \apj, 672, 765

\bibitem[{{Grazian} {et~al.}(2012){Grazian}, {Castellano}, {Fontana},
  {Pentericci}, {Dunlop}, {McLure}, {Koekemoer}, {Dickinson}, {Faber},
  {Ferguson}, {Galametz}, {Giavalisco}, {Grogin}, {Hathi}, {Kocevski}, {Lai},
  {Newman}, \& {Vanzella}}]{grazian12}
{Grazian}, A., {Castellano}, M., {Fontana}, A., et al. 2012, \aap, 547, A51

\bibitem[{{Grimes} {et~al.}(2009){Grimes}, {Heckman}, {Aloisi}, {Calzetti},
  {Leitherer}, {Martin}, {Meurer}, {Sembach}, \& {Strickland}}]{grimes09}
{Grimes}, J.~P., {Heckman}, T., {Aloisi}, A., et al. 2009,
  \apjs, 181, 272

\bibitem[{{Hayes} {et~al.}(2011){Hayes}, {Schaerer}, {{\"O}stlin}, {Mas-Hesse},
  {Atek}, \& {Kunth}}]{hayes11}
{Hayes}, M., {Schaerer}, D., {{\"O}stlin}, G., et al. 2011, \apj, 730, 8

\bibitem[{{Heckman} {et~al.}(2011){Heckman}, {Borthakur}, {Overzier},
  {Kauffmann}, {Basu-Zych}, {Leitherer}, {Sembach}, {Martin}, {Rich},
  {Schiminovich}, \& {Seibert}}]{heckman11}
{Heckman}, T.~M., {Borthakur}, S., {Overzier}, R., et al. 2011, \apj, 730, 5

\bibitem[{{Hurwitz} {et~al.}(1997){Hurwitz}, {Jelinsky}, \&
  {Dixon}}]{hurwitz97}
{Hurwitz}, M., {Jelinsky}, P., \& {Dixon}, W.~V.~D. 1997, \apjl, 481, L31

\bibitem[{{Inoue} {et~al.}(2014){Inoue}, {Shimizu}, {Iwata}, \&
  {Tanaka}}]{inoue14} {Inoue}, A.~K., {Shimizu}, I., {Iwata}, I., \&
{Tanaka}, M. 2014, \mnras, 442, 1805
	

\bibitem[{{Inoue} {et~al.}(2006){Inoue}, {Iwata}, \& {Deharveng}}]{inoue06}
{Inoue}, A.~K., {Iwata}, I., \& {Deharveng}, J.-M. 2006, \mnras, 371, L1

\bibitem[{{Iwata} {et~al.}(2009){Iwata}, {Inoue}, {Matsuda}, {Furusawa},
  {Hayashino}, {Kousai}, {Akiyama}, {Yamada}, {Burgarella}, \&
  {Deharveng}}]{iwata09}
{Iwata}, I., {Inoue}, A.~K., {Matsuda}, Y., et al. 2009, \apj, 692, 1287

\bibitem[{{Izotov} {et~al.}(2016a){Izotov}, {Orlitov{\'a}}, {Schaerer}, {Thuan},
  {Verhamme}, {Guseva}, \& {Worseck}}]{izotov16a}
{Izotov}, Y.~I., {Orlitov{\'a}}, I., {Schaerer}, D., et al. 2016a, \nat, 529, 178

\bibitem[{{Izotov} {et~al.}(2016b){Izotov}, {Schaerer}, {Thuan}, {Worseck}, {Guseva}, {Orlitov{\'a}}, \& {Verhamme}}]{izotov16b}
{Izotov}, Y.~I., {Schaerer}, D., {Thuan}, T.~X., et al. 2016b, arXiv, 1605.05160

\bibitem[{{Jones} {et~al.}(2013){Jones}, {Ellis}, {Schenker}, \&
  {Stark}}]{jones13}
{Jones}, T.~A., {Ellis}, R.~S., {Schenker}, M.~A., \& {Stark}, D.~P. 2013,
  \apj, 779, 52

\bibitem[{{Kennicutt}(1998)}]{kennicutt98}
{Kennicutt}, R.~C. 1998, \araa, 36, 189

\bibitem[{{Kornei} {et~al.}(2010){Kornei}, {Shapley}, {Erb}, {Steidel},
  {Reddy}, {Pettini}, \& {Bogosavljevi{\'c}}}]{kornei10}
{Kornei}, K.~A., {Shapley}, A.~E., {Erb}, D.~K., et al. 2010, \apj, 711, 693

\bibitem[{{Kroupa}(2001)}]{kroupa01}
{Kroupa}, P. 2001, \mnras, 322, 231

\bibitem[{{Kunth} {et~al.}(1998){Kunth}, {Mas-Hesse}, {Terlevich}, {Terlevich},
  {Lequeux}, \& {Fall}}]{kunth98}
{Kunth}, D., {Mas-Hesse}, J.~M., {Terlevich}, E., et al. 1998, \aap, 334, 11

\bibitem[{{Lehnert} \& {Heckman}(1995)}]{lehnert95}
{Lehnert}, M.~D. \& {Heckman}, T.~M. 1995, \apjs, 97, 89

\bibitem[{{Leitet} {et~al.}(2013){Leitet}, {Bergvall}, {Hayes}, {Linn{\'e}}, \&
  {Zackrisson}}]{leitet13}
{Leitet}, E., {Bergvall}, N., {Hayes}, M., {Linn{\'e}}, S., \& {Zackrisson}, E.
  2013, \aap, 553, A106

\bibitem[{{Leitherer} {et~al.}(2014){Leitherer}, {Ekstr{\"o}m}, {Meynet},
  {Schaerer}, {Agienko}, \& {Levesque}}]{leitherer14}
{Leitherer}, C., {Ekstr{\"o}m}, S., {Meynet}, G., et al. 2014, \apjs, 212, 14

\bibitem[{{Leitherer} {et~al.}(1995){Leitherer}, {Ferguson}, {Heckman}, \&
  {Lowenthal}}]{leitherer95b}
{Leitherer}, C., {Ferguson}, H.~C., {Heckman}, T.~M., \& {Lowenthal}, J.~D.
  1995, \apjl, 454, L19

\bibitem[{{Leitherer} {et~al.}(2016){Leitherer}, {Hernandez}, {Lee},
    \& {Oey}}]{leitherer16} {Leitherer}, C., {Hernandez}, S., {Lee},
  J.~C., \& {Oey}, M.~S. 2016, \apj, 823, 64

\bibitem[{{Leitherer} {et~al.}(2010){Leitherer}, {Ortiz Ot{\'a}lvaro},
    {Bresolin}, {Kudritzki}, {Lo Faro}, {Pauldrach}, {Pettini}, \&
    {Rix}}]{leitherer10} {Leitherer}, C., {Ortiz Ot{\'a}lvaro}, P.~A.,
  {Bresolin}, F., et al. 2010, \apjs, 189, 309

\bibitem[{{Leitherer} {et~al.}(1999){Leitherer}, {Schaerer},
    {Goldader}, {Delgado}, {Robert}, {Kune}, {de Mello}, {Devost}, \&
    {Heckman}}]{leitherer99} {Leitherer}, C., {Schaerer}, D.,
  {Goldader}, J.~D., et al. 1999, \apjs, 123, 3

\bibitem[{{Levesque} {et~al.}(2012){Levesque}, {Leitherer}, {Ekstrom},
  {Meynet}, \& {Schaerer}}]{levesque12}
{Levesque}, E.~M., {Leitherer}, C., {Ekstrom}, S., {Meynet}, G., \& {Schaerer},
  D. 2012, \apj, 751, 67

\bibitem[{{Lynds} \& {Sandage}(1963)}]{lynds63}
{Lynds}, C.~R. \& {Sandage}, A.~R. 1963, \apj, 137, 1005

\bibitem[{{Ma} {et~al.}(2016){Ma}, {Hopkins}, {Kasen}, {Quataert},
  {Faucher-Giguere}, {Keres}, \& {Murray}}]{ma16}
{Ma}, X., {Hopkins}, P.~F., {Kasen}, D., et al. 2016, \mnras, 459, 3614

\bibitem[{{Madau} \& {Haardt}(2015)}]{madau15}
{Madau}, P. \& {Haardt}, F. 2015, \apjl, 813, L8

\bibitem[{{Madau} {et~al.}(1999){Madau}, {Haardt}, \& {Rees}}]{madau99}
{Madau}, P., {Haardt}, F., \& {Rees}, M.~J. 1999, \apj, 514, 648

\bibitem[{{McDonald} \& {Miralda-Escud{\'e}}(2001)}]{mcdonald01}
{McDonald}, P. \& {Miralda-Escud{\'e}}, J. 2001, \apjl, 549, L11

\bibitem[{{McLure} {et~al.}(2010){McLure}, {Dunlop}, {Cirasuolo}, {Koekemoer},
  {Sabbi}, {Stark}, {Targett}, \& {Ellis}}]{mclure10}
{McLure}, R.~J., {Dunlop}, J.~S., {Cirasuolo}, M., et al. 2010, \mnras, 403,
  960

\bibitem[{{Mostardi} {et~al.}(2013){Mostardi}, {Shapley}, {Nestor}, {Steidel},
  {Reddy}, \& {Trainor}}]{mostardi13}
{Mostardi}, R.~E., {Shapley}, A.~E., {Nestor}, D.~B., et al. 2013, \apj, 779, 65

\bibitem[{{Mostardi} {et~al.}(2015){Mostardi}, {Shapley}, {Steidel}, {Trainor},
  {Reddy}, \& {Siana}}]{mostardi15}
{Mostardi}, R.~E., {Shapley}, A.~E., {Steidel}, C.~C., et al. 2015, \apj, 810, 107

\bibitem[{{Nestor} {et~al.}(2013){Nestor}, {Shapley}, {Kornei}, {Steidel}, \&
  {Siana}}]{nestor13}
{Nestor}, D.~B., {Shapley}, A.~E., {Kornei}, K.~A., {Steidel}, C.~C., \&
  {Siana}, B. 2013, \apj, 765, 47

\bibitem[{{Nestor} {et~al.}(2011){Nestor}, {Shapley}, {Steidel}, \&
  {Siana}}]{nestor11}
{Nestor}, D.~B., {Shapley}, A.~E., {Steidel}, C.~C., \& {Siana}, B. 2011, \apj,
  736, 18

\bibitem[{{Oesch} {et~al.}(2010){Oesch}, {Bouwens}, {Illingworth}, {Carollo},
  {Franx}, {Labb{\'e}}, {Magee}, {Stiavelli}, {Trenti}, \& {van
  Dokkum}}]{oesch10}
{Oesch}, P.~A., {Bouwens}, R.~J., {Illingworth}, G.~D., et al. 2010, \apjl, 709, L16

\bibitem[{{Oesch} {et~al.}(2013){Oesch}, {Bouwens}, {Illingworth}, {Labb{\'e}},
  {Franx}, {van Dokkum}, {Trenti}, {Stiavelli}, {Gonzalez}, \&
  {Magee}}]{oesch13b}
{Oesch}, P.~A., {Bouwens}, R.~J., {Illingworth}, G.~D., et al. 2013, \apj, 773, 75

\bibitem[{{Oey} {et~al.}(2002){Oey}, {Groves}, {Staveley-Smith}, \&
  {Smith}}]{oey02}
{Oey}, M.~S., {Groves}, B., {Staveley-Smith}, L., \& {Smith}, R.~C. 2002, \aj,
  123, 255

\bibitem[{{Oke} {et~al.}(1995){Oke}, {Cohen}, {Carr}, {Cromer}, {Dingizian},
  {Harris}, {Labrecque}, {Lucinio}, {Schaal}, {Epps}, \& {Miller}}]{oke95}
{Oke}, J.~B., {Cohen}, J.~G., {Carr}, M., et al. 1995, \pasp, 107, 375

\bibitem[{{O'Meara} {et~al.}(2013){O'Meara}, {Prochaska}, {Worseck},
    {Chen}, \& {Madau}}]{omeara13} {O'Meara}, J.~M., {Prochaska},
  J.~X., {Worseck}, G., {Chen}, H.-W., \& {Madau}, P. 2013, \apj, 765,
  137

\bibitem[{{Putman} {et~al.}(2012){Putman}, {Peek}, \& {Joung}}]{putman12}
{Putman}, M.~E., {Peek}, J.~E.~G., \& {Joung}, M.~R. 2012, \araa, 50, 491

\bibitem[{{Razoumov} \& {Sommer-Larsen}(2010)}]{razoumov10}
{Razoumov}, A.~O. \& {Sommer-Larsen}, J. 2010, \apj, 710, 1239

\bibitem[{{Reddy} {et~al.}(2012){Reddy}, {Dickinson}, {Elbaz}, {Morrison},
  {Giavalisco}, {Ivison}, {Papovich}, {Scott}, {Buat}, {Burgarella},
  {Charmandaris}, {Daddi}, {Magdis}, {Murphy}, {Altieri}, {Aussel},
  {Dannerbauer}, {Dasyra}, {Hwang}, {Kartaltepe}, {Leiton}, {Magnelli}, \&
  {Popesso}}]{reddy12a}
{Reddy}, N., {Dickinson}, M., {Elbaz}, D., et al. 2012, \apj, 744, 154

\bibitem[{{Reddy} {et~al.}(2015){Reddy}, {Kriek}, {Shapley}, {Freeman},
  {Siana}, {Coil}, {Mobasher}, {Price}, {Sanders}, \& {Shivaei}}]{reddy15}
{Reddy}, N.~A., {Kriek}, M., {Shapley}, A.~E., et al. 2015, \apj, 806, 259

\bibitem[{{Reddy} \& {Steidel}(2009)}]{reddy09}
{Reddy}, N.~A. \& {Steidel}, C.~C. 2009, \apj, 692, 778

\bibitem[{{Reddy} {et~al.}(2008){Reddy}, {Steidel}, {Pettini}, {Adelberger},
  {Shapley}, {Erb}, \& {Dickinson}}]{reddy08}
{Reddy}, N.~A., {Steidel}, C.~C., {Pettini}, M., et al. 2008, \apjs, 175, 48

\bibitem[{{Rivera-Thorsen} {et~al.}(2015){Rivera-Thorsen}, {Hayes},
  {{\"O}stlin}, {Duval}, {Orlitov{\'a}}, {Verhamme}, {Mas-Hesse}, {Schaerer},
  {Cannon}, {Ot{\'{\i}}-Floranes}, {Sandberg}, {Guaita}, {Adamo}, {Atek},
  {Herenz}, {Kunth}, {Laursen}, \& {Melinder}}]{rivera15}
{Rivera-Thorsen}, T.~E., {Hayes}, M., {{\"O}stlin}, G., et al. 2015, \apj, 805, 14

\bibitem[{{Rix} {et~al.}(2004){Rix}, {Pettini}, {Leitherer}, {Bresolin},
  {Kudritzki}, \& {Steidel}}]{rix04}
{Rix}, S.~A., {Pettini}, M., {Leitherer}, C., et al. 2004, \apj, 615, 98

\bibitem[{{Robertson} {et~al.}(2010){Robertson}, {Ellis}, {Dunlop}, {McLure},
  \& {Stark}}]{brant10}
{Robertson}, B.~E., {Ellis}, R.~S., {Dunlop}, J.~S., {McLure}, R.~J., \&
  {Stark}, D.~P. 2010, \nat, 468, 49

\bibitem[{{Robertson} {et~al.}(2015){Robertson}, {Ellis}, {Furlanetto}, \&
  {Dunlop}}]{brant15}
{Robertson}, B.~E., {Ellis}, R.~S., {Furlanetto}, S.~R., \& {Dunlop}, J.~S.
  2015, \apjl, 802, L19

\bibitem[{{Rudie} {et~al.}(2013){Rudie}, {Steidel}, {Shapley}, \&
  {Pettini}}]{rudie13}
{Rudie}, G.~C., {Steidel}, C.~C., {Shapley}, A.~E., \& {Pettini}, M. 2013,
  \apj, 769, 146

\bibitem[{{Salpeter}(1955)}]{salpeter55}
{Salpeter}, E.~E. 1955, \apj, 121, 161

\bibitem[{{Shapiro} \& {Field}(1976)}]{shapiro76}
{Shapiro}, P.~R. \& {Field}, G.~B. 1976, \apj, 205, 762

\bibitem[{{Shapley} {et~al.}(2003){Shapley}, {Steidel}, {Pettini}, \&
  {Adelberger}}]{shapley03}
{Shapley}, A.~E., {Steidel}, C.~C., {Pettini}, M., \& {Adelberger}, K.~L. 2003,
  \apj, 588, 65

\bibitem[{{Shapley} {et~al.}(2006){Shapley}, {Steidel}, {Pettini},
  {Adelberger}, \& {Erb}}]{shapley06}
{Shapley}, A.~E., {Steidel}, C.~C., {Pettini}, M., {Adelberger}, K.~L., \&
  {Erb}, D.~K. 2006, \apj, 651, 688

\bibitem[{{Shapley} {et~al.}(2016){Shapley}, {Steidel}, {Strom},
    {Bogosavljevi{\'c}}, {Reddy}, {Siana}, {Mostardi}, \& {Rudie}}]{shapley16}
{Shapley}, A.~E., {Steidel}, C.~C., {Strom}, A.~L., et al. 2016, 
arXiv, 1606.00443

\bibitem[{{Siana} {et~al.}(2015){Siana}, {Shapley}, {Kulas}, {Nestor},
  {Steidel}, {Teplitz}, {Alavi}, {Brown}, {Conselice}, {Ferguson}, {Dickinson},
  {Giavalisco}, {Colbert}, {Bridge}, {Gardner}, \& {de Mello}}]{siana15}
{Siana}, B., {Shapley}, A.~E., {Kulas}, K.~R., et al. 2015, \apj, 804, 17

\bibitem[{{Siana} {et~al.}(2008){Siana}, {Teplitz}, {Chary}, {Colbert}, \&
  {Frayer}}]{siana08}
{Siana}, B., {Teplitz}, H.~I., {Chary}, R.-R., {Colbert}, J., \& {Frayer},
  D.~T. 2008, \apj, 689, 59

\bibitem[{{Siana} {et~al.}(2007){Siana}, {Teplitz}, {Colbert}, {Ferguson},
  {Dickinson}, {Brown}, {Conselice}, {de Mello}, {Gardner}, {Giavalisco}, \&
  {Menanteau}}]{siana07}
{Siana}, B., {Teplitz}, H.~I., {Colbert}, J., et al. 2007, \apj, 668, 62

\bibitem[{{Siana} {et~al.}(2010){Siana}, {Teplitz}, {Ferguson}, {Brown},
  {Giavalisco}, {Dickinson}, {Chary}, {de Mello}, {Conselice}, {Bridge},
  {Gardner}, {Colbert}, \& {Scarlata}}]{siana10}
{Siana}, B., {Teplitz}, H.~I., {Ferguson}, H.~C., et al. 2010,
  \apj, 723, 241

\bibitem[{{Silk}(1997)}]{silk97}
{Silk}, J. 1997, \apj, 481, 703

\bibitem[{{Stanway} {et~al.}(2016){Stanway}, {Eldridge}, \&
  {Becker}}]{stanway16}
{Stanway}, E.~R., {Eldridge}, J.~J., \& {Becker}, G.~D. 2016, \mnras, 456, 485

\bibitem[{{Stark} {et~al.}(2007){Stark}, {Ellis}, {Richard}, {Kneib}, {Smith},
  \& {Santos}}]{stark07}
{Stark}, D.~P., {Ellis}, R.~S., {Richard}, J., et al. 2007, \apj, 663, 10

\bibitem[{{Steidel} {et~al.}(1999){Steidel}, {Adelberger}, {Giavalisco},
  {Dickinson}, \& {Pettini}}]{steidel99}
{Steidel}, C.~C., {Adelberger}, K.~L., {Giavalisco}, M., {Dickinson}, M., \&
  {Pettini}, M. 1999, \apj, 519, 1

\bibitem[{{Steidel} {et~al.}(2003){Steidel}, {Adelberger}, {Shapley},
  {Pettini}, {Dickinson}, \& {Giavalisco}}]{steidel03}
{Steidel}, C.~C., {Adelberger}, K.~L., {Shapley}, A.~E., et al. 2003, \apj, 592, 728

\bibitem[{{Steidel} {et~al.}(2011){Steidel}, {Bogosavljevi{\'c}}, {Shapley},
  {Kollmeier}, {Reddy}, {Erb}, \& {Pettini}}]{steidel11}
{Steidel}, C.~C., {Bogosavljevi{\'c}}, M., {Shapley}, A.~E., et al. 2011, \apj, 736, 160

\bibitem[{{Steidel} {et~al.}(2010){Steidel}, {Erb}, {Shapley}, {Pettini},
  {Reddy}, {Bogosavljevi{\'c}}, {Rudie}, \& {Rakic}}]{steidel10}
{Steidel}, C.~C., {Erb}, D.~K., {Shapley}, A.~E., et al. 2010, \apj, 717, 289

\bibitem[{{Steidel} {et~al.}(2001){Steidel}, {Pettini}, \&
  {Adelberger}}]{steidel01}
{Steidel}, C.~C., {Pettini}, M., \& {Adelberger}, K.~L. 2001, \apj, 546, 665

\bibitem[{{Steidel} {et~al.}(2004){Steidel}, {Shapley}, {Pettini},
  {Adelberger}, {Erb}, {Reddy}, \& {Hunt}}]{steidel04}
{Steidel}, C.~C., {Shapley}, A.~E., {Pettini}, M., et al. 2004, \apj, 604, 534

\bibitem[{{Steidel} {et~al.}(2016){Steidel}, {Strom}, {Pettini}, {Rudie},
{Reddy}, {Trainor}}]{steidel16}
{Steidel}, C.~C., {Strom}, A.~L., {Pettini}, M., et al. 2016, arXiv, 1605.07186

\bibitem[{{Strickland} {et~al.}(2004){Strickland}, {Heckman}, {Colbert},
  {Hoopes}, \& {Weaver}}]{strickland04b}
{Strickland}, D.~K., {Heckman}, T.~M., {Colbert}, E.~J.~M., {Hoopes}, C.~G., \&
  {Weaver}, K.~A. 2004, \apj, 606, 829

\bibitem[{{Topping} \& {Shull}(2015)}]{topping15}
{Topping}, M.~W. \& {Shull}, J.~M. 2015, \apj, 800, 97

\bibitem[{{Trainor} {et~al.}(2015){Trainor}, {Steidel}, {Strom}, \&
  {Rudie}}]{trainor15}
{Trainor}, R.~F., {Steidel}, C.~C., {Strom}, A.~L., \& {Rudie}, G.~C. 2015,
  \apj, 809, 89

\bibitem[{{Vanzella} {et~al.}(2010{\natexlab{a}}){Vanzella}, {Giavalisco},
  {Inoue}, {Nonino}, {Fontanot}, {Cristiani}, {Grazian}, {Dickinson}, {Stern},
  {Tozzi}, {Giallongo}, {Ferguson}, {Spinrad}, {Boutsia}, {Fontana}, {Rosati},
  \& {Pentericci}}]{vanzella10b}
{Vanzella}, E., {Giavalisco}, M., {Inoue}, A.~K., et al. 2010{\natexlab{a}}, \apj, 725, 1011

\bibitem[{{Vanzella} {et~al.}(2012){Vanzella}, {Guo}, {Giavalisco}, {Grazian},
  {Castellano}, {Cristiani}, {Dickinson}, {Fontana}, {Nonino}, {Giallongo},
  {Pentericci}, {Galametz}, {Faber}, {Ferguson}, {Grogin}, {Koekemoer},
  {Newman}, \& {Siana}}]{vanzella12}
{Vanzella}, E., {Guo}, Y., {Giavalisco}, M., et al. 2012,
  \apj, 751, 70

\bibitem[{{Vanzella} {et~al.}(2010{\natexlab{b}}){Vanzella}, {Siana},
  {Cristiani}, \& {Nonino}}]{vanzella10a}
{Vanzella}, E., {Siana}, B., {Cristiani}, S., \& {Nonino}, M.
  2010{\natexlab{b}}, \mnras, 404, 1672

\bibitem[{{Welty} {et~al.}(2012){Welty}, {Xue}, \& {Wong}}]{welty12}
{Welty}, D.~E., {Xue}, R., \& {Wong}, T. 2012, \apj, 745, 173

\bibitem[{{Wofford} {et~al.}(2013){Wofford}, {Leitherer}, \&
  {Salzer}}]{wofford13}
{Wofford}, A., {Leitherer}, C., \& {Salzer}, J. 2013, \apj, 765, 118

\bibitem[{{Yan} \& {Windhorst}(2004)}]{yan04}
{Yan}, H. \& {Windhorst}, R.~A. 2004, \apjl, 600, L1

\bibitem[{{Yoon} \& {Langer}(2005)}]{yoon05}
{Yoon}, S.-C. \& {Langer}, N. 2005, \aap, 443, 643

\bibitem[{{Zackrisson} {et~al.}(2013){Zackrisson}, {Inoue}, \&
  {Jensen}}]{zackrisson13}
{Zackrisson}, E., {Inoue}, A.~K., \& {Jensen}, H. 2013, \apj, 777, 39

\end{thebibliography}








\end{document}